\documentclass[final,journal,twocolumn]{IEEEtran}
\usepackage{cite}
\usepackage{url}
\usepackage{lipsum}
\usepackage{siunitx}
\usepackage{pifont}
\usepackage[pdftex]{graphicx}
\usepackage{amsmath}
\usepackage{upgreek}
\usepackage{amsfonts}
\usepackage{makecell}
\usepackage{multirow}
\usepackage{booktabs}
\usepackage{tabularx}
\usepackage[usenames,dvipsnames]{xcolor} 
\usepackage{colortbl}
\usepackage{pgffor}
\usepackage[export]{adjustbox}
\usepackage[short]{optidef}
\usepackage{soul}
\usepackage{ifthen}
\usepackage{dsfont}
\usepackage{amssymb}

\graphicspath{../figures/}
\newcommand{\cmark}{\textcolor{myblue}{\ding{51}}} 

\usepackage{tikz}
\usepackage{pgfplots}
\usepackage{pgfplotstable}
\usepackage{mathtools}
\usepackage{color}
\usepackage{xcolor}

\usepgfplotslibrary{colorbrewer}
\pgfplotsset{cycle list/Set1-9}
\tikzset{every picture/.style={line width=1pt}}
\usetikzlibrary{patterns,backgrounds}
\usepgfplotslibrary{fillbetween}
\pgfplotsset{
  tick label style = {font=\sansmath\sffamily\footnotesize},
  every axis label = {font=\sansmath\sffamily\footnotesize},
  y label style={at={(0.05,0.5)}},
  legend style = {font=\sansmath\sffamily\scriptsize},
  label style = {font=\sansmath\sffamily\footnotesize}
}
\pgfplotsset{compat=1.3} %
\DeclareMathAlphabet\mathbfcal{OMS}{cmsy}{b}{n}

\usepackage[most]{tcolorbox}
\tcbset{top=0.01mm,left=0.01mm,bottom=0.01mm,right=0.01mm}
\usepackage{efbox}
\efboxsetup{linecolor=green,linewidth=10pt}

\usepackage{booktabs,colortbl}

\usepackage{pdftexcmds}

\usepackage{xspace}

\newcommand{\basemae}{CSMAE\xspace}

\def\compilefigs{0}
\newcommand{\inputfig}[1]{\if\compilefigs1\input{sources/plots/#1}\else\includegraphics{figures/#1}\fi}
\newcommand{\inputtable}[1]{\if\compilefigs1\input{sources/plots/#1}\else\includegraphics{figures/#1}\fi}

\def\cmark{\ding{51}} 
\def\xmark{\ding{55}}

\begin{document}

\title{Exploring Masked Autoencoders for Sensor-Agnostic Image Retrieval in Remote Sensing}

\author{Jakob~Hackstein,~Gencer~Sumbul,~\IEEEmembership{Member,~IEEE},~Kai~Norman~Clasen,~\IEEEmembership{Member,~IEEE},~and Beg{\"u}m~Demir,~\IEEEmembership{Senior~Member,~IEEE}
\thanks{Jakob Hackstein, Kai Norman Clasen and Beg{\"u}m Demir are with the Faculty of Electrical Engineering and Computer Science, Technische Universit\"at Berlin, 10623 Berlin, Germany, also with the BIFOLD - Berlin Institute for the Foundations of Learning and Data, 10623 Berlin, Germany (e-mail: \mbox{hackstein@campus.tu-berlin.de}; \mbox{k.clasen@tu-berlin.de}; \mbox{demir@tu-berlin.de}).

Gencer Sumbul is with the Environmental Computational Science and Earth Observation Laboratory (ECEO), École Polytechnique Fédérale de Lausanne (EPFL), 1950 Sion, Switzerland (e-mail: \mbox{gencer.sumbul@epfl.ch}).}%
}

\maketitle

\begin{abstract}
Self-supervised learning through masked autoencoders (MAEs) has recently attracted great attention for remote sensing (RS) image representation learning, and thus embodies a significant potential for content-based image retrieval (CBIR) from ever-growing RS image archives. However, the existing MAE based CBIR studies in RS assume that the considered RS images are acquired by a single image sensor, and thus are only suitable for uni-modal CBIR problems. The effectiveness of MAEs for cross-sensor CBIR, which aims to search semantically similar images across different image modalities, has not been explored yet. In this paper, we take the first step to explore the effectiveness of MAEs for sensor-agnostic CBIR in RS. To this end, we present a systematic overview on the possible adaptations of the vanilla MAE to exploit masked image modeling on multi-sensor RS image archives (denoted as cross-sensor masked autoencoders [\basemae{s}]) in the context of CBIR. Based on different adjustments applied to the vanilla MAE, we introduce different~\basemae~models. We also provide an extensive experimental analysis of these~\basemae~models. We finally derive a guideline to exploit masked image modeling for uni-modal and cross-modal CBIR problems in RS. The code of this work is publicly available at \url{https://github.com/jakhac/CSMAE}.
\end{abstract}


\section{Introduction}
Content-based image retrieval (CBIR) from large-scale remote sensing (RS) image archives is one of the most important research topics in RS \mbox{\cite{Tang:2022}}. CBIR aims to search for images similar to a given query image based on their semantic content. Thus, the accurate characterization of RS image representations is of great importance for CBIR~\cite{Sumbul:2022:2}. To this end, deep learning (DL) based image representation learning (IRL) has been found successful in RS \mbox{\cite{Chen:2023, Su:2023}}. For a detailed review of DL-based CBIR studies in RS, we refer readers to~\cite{Sumbul:2021, Sumbul:2022:2}. Compared to supervised DL-based IRL methods, self-supervised IRL has recently attracted great attention for remote sensing (RS) images. Since it does not require the availability of training images annotated by land-use land-cover class labels (which can be time-consuming and expensive due to labeling costs), it allows to leverage ever-growing RS image archives for learning the parameters of DL models on a vast amount of RS images. Contrastive learning-based methods have initially paved the way for employing self-supervised IRL on RS images (e.g., \cite{Sumbul:2022, Jung:2022}). Recently, masked image modeling (MIM)~\cite{Zhenda:2022, He:2022} (i.e., self-supervised learning through masked autoencoders [MAEs]) has been also found successful in RS~\cite{Wang:2023:2, Dilxat:2023, Keumgang:2023, Reed:2023, Wang:2023, Xian:2023} (see Section \ref{sec:related_work}) for self-supervised IRL. MAEs characterize image representations based on the pretext task of reconstructing images from their masked patches by dynamizing vision transformers (ViTs)~\cite{Dosovitskiy:2021}. Their success on IRL relies on two main reasons. First, in contrast to contrastive self-supervised learning, MAEs are capable of accurately learning image representations without the application of any data augmentation strategies. This is of particular importance for RS images since most of the data augmentation strategies are designed for natural images and their direct adaptation to RS images may not be always feasible. Second, it has been shown that MAEs in combination with ViTs can be effectively scaled into larger DL models in proportion to the amount of training data~\cite{He:2022, Li:2022}. 

Although the existing studies have proven the success of MAEs for accurately learning RS image representations (see Section \ref{sec:related_work}), existing MIM approaches developed for CBIR in RS are based on RS images acquired by a single image sensor. Thus, they are suitable to be utilized for only uni-modal CBIR problems. However, in RS, a CBIR task can be formulated as searching for semantically similar images across different image modalities that is known as cross-sensor CBIR~\cite{Yansheng:2018, Xiong:2020, Chaudhuri:2020, Sun:2022, Chaudhuri:2022, Sumbul:2022} (see Section \ref{sec:related_work}). When IRL is employed for cross-sensor CBIR, it is crucial to model not only intra-modal similarities, but also inter-modal similarities. This requires to learn RS image representations from multi-modal RS image archives. The direct adaptation of MAEs to model inter-modal similarities across multi-modal RS image archives is challenging, since MAEs are originally designed to reconstruct uni-modal images. To address this issue, as a first time in RS, we explore the effectiveness of MAEs for sensor-agnostic image retrieval in RS. The main contributions of this paper are summarized as follows:
\begin{itemize}
    \item We present a systematic overview on the possible adaptations of the vanilla MAE~\cite{He:2022} to exploit MIM in the context of cross-modal CBIR based on cross-sensor masked autoencoders (\basemae{s}).
    \item We introduce different~\basemae~models based on: i) the architectural adjustments applied to the considered ViTs; ii) the adaptations on image masking; and iii) the reformulation of masked image modeling. 
    \item We provide extensive experimental results of the introduced~\basemae~models, while conducting sensitivity analysis, ablation study and comparison with other approaches.
    \item We derive some guidelines to utilize MIM for uni-modal and cross-modal RS image retrieval problems.
\end{itemize}
The remaining part of this paper is organized as follows: Section \ref{sec:related_work} presents the related works on MAEs and cross-sensor CBIR in RS. In Section \ref{sec:method}, we introduce different~\basemae~models. Experimental analysis of \basemae{s} is carried out in Section~\ref{sec:result}, while Section~\ref{sec:exp_setup} provides the design of experiments. Section~\ref{sec:conclusion} concludes our paper with some guidelines for utilizing MAEs in the framework of sensor-agnostic RS image retrieval. 

\section{Related Work}
In this section, we initially present the recent advances on MAEs in RS and then survey the existing cross-sensor CBIR methods in RS.
\label{sec:related_work}
\subsection{Masked Autoencoders in RS}
In RS, MAEs are mainly employed for learning ViTs on a vast amount of uni-modal RS images (i.e., pre-trained model) in a self-supervised way~\cite{Gao:2022:2, Cong:2022, Wang:2023:2, Dilxat:2023, Keumgang:2023, Reed:2023, Wang:2023, Xian:2023}. Then, the pre-trained ViTs are fine-tuned on annotated RS images (which is typically on a smaller scale than pre-training in terms of data size) for various downstream tasks (e.g., semantic segmentation, scene classification etc.). As an example, in \cite{Gao:2022:2, Wang:2023:2} pre-trained models based on the vanilla MAE are utilized for scene classification problems. In \cite{Keumgang:2023}, they are applied for object detection and semantic segmentation problems, while the considered ViTs are scaled up by connecting the multi-head self-attention
structures and feed-forward blocks in parallel. In \cite{Dilxat:2023}, contrastive learning is combined with masked image modeling in a self-distillation way, aiming to achieve both global semantic separability and local spatial perceptibility. In this study, the vanilla MAE is employed to also reconstruct RS images in the frequency
domain for change detection problems. In \cite{Reed:2023}, the learning objective of the vanilla MAE is transformed into the reconstruction of high-frequency and low-frequency RS image features, while a ground sample distance positional encoding strategy is utilized. In \cite{Wang:2023}, rotated varied-size window attention is replaced with the full attention in the ViTs of vanilla MAE, aiming to improve object representation learning on RS images and to reduce the computational cost of the considered ViTs. In \cite{Xian:2023}, image masking in the vanilla MAE is tailored towards the specifics of RS data by maintaining a small number of pixels in masked areas. This results in enhanced performance for various downstream tasks at the cost of scalability since training requires to encode complete images instead of unmasked parts only. In~\cite{Cong:2022}, to leverage temporal information of RS images, temporal embeddings are employed together with image masking across time.


In the context of foundation models, MAEs have emerged as a powerful framework for learning versatile representations. Recent research \mbox{\cite{xiong2024neural, han2024bridging}} explores the integration of multiple modalities combined with large-scale datasets to obtain richer representations that perform well on various downstream tasks. To this end, ViTs are modified to process several modalities, while the pre-training objective is extended accordingly. For instance, in~\mbox{\cite{xiong2024one, astruc2024omnisat}} it is proposed to employ individual patch embedding layers for each modality, which are then fed into a shared ViT backbone. In \mbox{\cite{irvin2023usat}} groups of multi-sensor data based on common ground sampling distance are processed individually, instead of resizing to a fixed resolution, upon applying a backbone shared among all sensors. This approach preserves information captured with high spatial resolution and allows specialized treatment of each sensor group. In \mbox{\cite{nedungadi2024mmearth}}, a multi-pretext task objective is formulated to capture unique characteristics of each modality in latent representations.  Note that in recent years, the foundation models have gained an increasing interest in RS. However, to the best of our knowledge, the analysis on architectural choices as well as optimization objectives that play an important role for an accurate IRL in the context of multi-modal MAEs have not been adequately studied yet.


\subsection{Cross-Sensor CBIR in RS}
Most of the existing cross-sensor CBIR methods in RS \mbox{\cite{Yansheng:2018, Xiong:2020, Chaudhuri:2020, Sun:2022}} assume the availability of multi-modal training images annotated by land-use land-cover class labels. As an example, in \mbox{\cite{Chaudhuri:2020}} a cross-sensor CBIR framework is proposed, aiming to learn an unified representation space over the sensor-specific RS image features. To this end, a two-stage training procedure is utilized. In the first stage, supervised sensor-specific feature learning is employed with the cross-entropy objective for intra-modal similarity preservation. In the second stage, the mean squared error, cross-entropy and reconstruction losses are utilized for learning the unified representation space across different image modalities. In \mbox{\cite{Xiong:2020}}, a deep cross-modality hashing network is introduced to model intra-modal RS image similarities by using the supervised triplet loss function over the representations of each RS image modality. In \mbox{\cite{Sun:2022}}, a deep hashing method is proposed for cross-sensor CBIR problems in RS. It utilizes multi-sensor fusion with synthetically generated images to mitigate inter-modal discrepancy among different RS image modalities, while exploiting supervised cross-entropy objective on each modality for intra-modal similarity preservation. 

\input{sources/method_figures/overall_model}
We would like to note that a sufficient number of annotated multi-modal training images may not be always available in RS. Moreover, collecting reliable land-use land-cover class labels can be time-consuming, complex and costly to gather \mbox{\cite{Sumbul:2022, Sumbul:2023}}. To address this issue, in \mbox{\cite{Chaudhuri:2022}} label-deficit zero-shot learning (LDZSL) training protocol is introduced. LDZSL aims to map the information learned by annotated training samples of the seen classes into unseen classes by facilitating Siamese neural networks with word2vec encodings of class labels. In \mbox{\cite{Sumbul:2022}}, a self-supervised cross-sensor CBIR method is introduced to model mutual information between different RS image modalities without any need for annotated training image. To this end, it employs contrastive learning of RS image representations by maximizing agreement between images acquired by different sensors on the same geographical area.

\section{Cross-Sensor Masked Autoencoders (\basemae) for Sensor-Agnostic Remote Sensing Image Retrieval}
\label{sec:method}
Let $\mathcal{X}^1$ and $\mathcal{X}^2$ be two image archives associated with a different image modality (acquired by a different sensor). Each archive $\mathcal{X}^j=\{\boldsymbol{x}_i^j\}_{i=1}^{N_j} \:\: \forall j\in\{1,2\}$ includes $N_j$ images, where $\boldsymbol{x}_i^j$ is the $i$th RS image in the $j$th image archive and $(\boldsymbol{x}_i^1, \boldsymbol{x}_i^2)$ is the $i$th multi-modal image pair that includes two images acquired by different sensors on the same geographical area. We assume that an unlabeled training set $\mathcal{T}=\{(\boldsymbol{x}_i^1, \boldsymbol{x}_i^2)\}_{i=1}^D$, which includes $D$ pairs of multi-modal images, is available. A sensor-agnostic CBIR problem can be formulated as finding images from the image archive $\mathcal{X}^j$ that are similar to a given query from the image archive $\mathcal{X}^k$ based on the similarities in image representation (i.e., feature) space. In the case of $j \neq k$ it becomes a cross-sensor CBIR task, while $j = k$ for an uni-modal CBIR task. For both retrieval tasks, the characterization of the semantic content of RS images on image representations is of great importance for accurate RS CBIR performance. For image representation learning, masked image modeling (MIM)~\cite{Zhenda:2022, He:2022} (i.e., self-supervised learning through masked autoencoders [MAEs]) has recently emerged as one of the most successful approaches. MAEs are originally designed for only uni-modal images (i.e., either $j=1$ or $j=2$), and thus can be utilized for only uni-modal CBIR problems. As a first time, in this paper, we explore the effectiveness of MAEs for sensor-agnostic CBIR problems. To this end, in the following subsections, we first provide general information on vanilla MAEs, and then present our adaptations on MAEs to exploit MIM for both uni-modal and cross-sensor CBIR problems. 

\begin{figure*}[t]
   \newcommand{\figwidth}{0.3\linewidth}
    \newcommand{\figheight}{1.15in}
    \centering
     \begin{minipage}[t]{\figwidth}
        \centering
        \centerline{

\tikzset{every picture/.style={line width=0.75pt}} 

\begin{tikzpicture}[x=0.75pt,y=0.75pt,yscale=-1,xscale=1]

\draw (35.25,35.25) node  {\includegraphics[width=52.5pt,height=52.5pt]{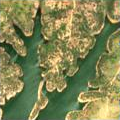}};
\draw (125.25,35.25) node  {\includegraphics[width=52.5pt,height=52.5pt]{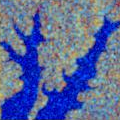}};
\draw  [draw opacity=0] (0.24,0.27) -- (70.73,0.23) -- (70.77,70.73) -- (0.28,70.77) -- cycle ; \draw   (0.24,0.27) -- (0.28,70.77)(14.24,0.26) -- (14.28,70.76)(28.24,0.25) -- (28.28,70.75)(42.24,0.25) -- (42.28,70.75)(56.24,0.24) -- (56.28,70.74)(70.24,0.23) -- (70.28,70.73) ; \draw   (0.24,0.27) -- (70.73,0.23)(0.25,14.27) -- (70.74,14.23)(0.25,28.27) -- (70.75,28.23)(0.26,42.27) -- (70.75,42.23)(0.27,56.27) -- (70.76,56.23)(0.28,70.27) -- (70.77,70.23) ; \draw    ;
\draw  [draw opacity=0] (90.25,0.29) -- (160.74,0.25) -- (160.78,70.75) -- (90.29,70.79) -- cycle ; \draw   (90.25,0.29) -- (90.29,70.79)(104.25,0.28) -- (104.29,70.78)(118.25,0.27) -- (118.29,70.77)(132.25,0.26) -- (132.29,70.76)(146.25,0.26) -- (146.29,70.76)(160.25,0.25) -- (160.29,70.75) ; \draw   (90.25,0.29) -- (160.74,0.25)(90.26,14.29) -- (160.75,14.25)(90.26,28.29) -- (160.76,28.25)(90.27,42.29) -- (160.76,42.25)(90.28,56.29) -- (160.77,56.25)(90.29,70.29) -- (160.78,70.25) ; \draw    ;
\draw  [draw opacity=0][fill={rgb, 255:red, 126; green, 211; blue, 33 }  ,fill opacity=1 ] (28.28,56.25) -- (42.53,56.25) -- (42.53,70.75) -- (28.28,70.75) -- cycle ; \draw   (28.28,56.25) -- (28.28,70.75)(42.28,56.25) -- (42.28,70.75) ; \draw   (28.28,56.25) -- (42.53,56.25)(28.28,70.25) -- (42.53,70.25) ; \draw    ;
\draw  [draw opacity=0][fill={rgb, 255:red, 126; green, 211; blue, 33 }  ,fill opacity=1 ] (56.27,42.23) -- (70.52,42.23) -- (70.52,56.73) -- (56.27,56.73) -- cycle ; \draw   (56.27,42.23) -- (56.27,56.73)(70.27,42.23) -- (70.27,56.73) ; \draw   (56.27,42.23) -- (70.52,42.23)(56.27,56.23) -- (70.52,56.23) ; \draw    ;
\draw  [draw opacity=0][fill={rgb, 255:red, 126; green, 211; blue, 33 }  ,fill opacity=1 ] (0.25,14.27) -- (14.5,14.27) -- (14.5,28.77) -- (0.25,28.77) -- cycle ; \draw   (0.25,14.27) -- (0.25,28.77)(14.25,14.27) -- (14.25,28.77) ; \draw   (0.25,14.27) -- (14.5,14.27)(0.25,28.27) -- (14.5,28.27) ; \draw    ;
\draw  [draw opacity=0][fill={rgb, 255:red, 126; green, 211; blue, 33 }  ,fill opacity=1 ] (42.25,0.25) -- (56.5,0.25) -- (56.5,14.75) -- (42.25,14.75) -- cycle ; \draw   (42.25,0.25) -- (42.25,14.75)(56.25,0.25) -- (56.25,14.75) ; \draw   (42.25,0.25) -- (56.5,0.25)(42.25,14.25) -- (56.5,14.25) ; \draw    ;
\draw  [draw opacity=0][fill={rgb, 255:red, 126; green, 211; blue, 33 }  ,fill opacity=1 ] (28.25,28.25) -- (42.5,28.25) -- (42.5,42.75) -- (28.25,42.75) -- cycle ; \draw   (28.25,28.25) -- (28.25,42.75)(42.25,28.25) -- (42.25,42.75) ; \draw   (28.25,28.25) -- (42.5,28.25)(28.25,42.25) -- (42.5,42.25) ; \draw    ;
\draw  [draw opacity=0][fill={rgb, 255:red, 126; green, 211; blue, 33 }  ,fill opacity=1 ] (0.26,42.27) -- (14.51,42.27) -- (14.51,56.77) -- (0.26,56.77) -- cycle ; \draw   (0.26,42.27) -- (0.26,56.77)(14.26,42.27) -- (14.26,56.77) ; \draw   (0.26,42.27) -- (14.51,42.27)(0.26,56.27) -- (14.51,56.27) ; \draw    ;
\draw  [draw opacity=0][fill={rgb, 255:red, 126; green, 211; blue, 33 }  ,fill opacity=1 ] (132.25,0.26) -- (146.5,0.26) -- (146.5,14.76) -- (132.25,14.76) -- cycle ; \draw   (132.25,0.26) -- (132.25,14.76)(146.25,0.26) -- (146.25,14.76) ; \draw   (132.25,0.26) -- (146.5,0.26)(132.25,14.26) -- (146.5,14.26) ; \draw    ;
\draw  [draw opacity=0][fill={rgb, 255:red, 126; green, 211; blue, 33 }  ,fill opacity=1 ] (118.29,56.26) -- (132.54,56.26) -- (132.54,70.76) -- (118.29,70.76) -- cycle ; \draw   (118.29,56.26) -- (118.29,70.76)(132.29,56.26) -- (132.29,70.76) ; \draw   (118.29,56.26) -- (132.54,56.26)(118.29,70.26) -- (132.54,70.26) ; \draw    ;
\draw  [draw opacity=0][fill={rgb, 255:red, 126; green, 211; blue, 33 }  ,fill opacity=1 ] (118.27,28.27) -- (132.52,28.27) -- (132.52,42.77) -- (118.27,42.77) -- cycle ; \draw   (118.27,28.27) -- (118.27,42.77)(132.27,28.27) -- (132.27,42.77) ; \draw   (118.27,28.27) -- (132.52,28.27)(118.27,42.27) -- (132.52,42.27) ; \draw    ;
\draw  [draw opacity=0][fill={rgb, 255:red, 126; green, 211; blue, 33 }  ,fill opacity=1 ] (90.26,14.29) -- (104.51,14.29) -- (104.51,28.79) -- (90.26,28.79) -- cycle ; \draw   (90.26,14.29) -- (90.26,28.79)(104.26,14.29) -- (104.26,28.79) ; \draw   (90.26,14.29) -- (104.51,14.29)(90.26,28.29) -- (104.51,28.29) ; \draw    ;
\draw  [draw opacity=0][fill={rgb, 255:red, 126; green, 211; blue, 33 }  ,fill opacity=1 ] (90.27,42.28) -- (104.52,42.28) -- (104.52,56.78) -- (90.27,56.78) -- cycle ; \draw   (90.27,42.28) -- (90.27,56.78)(104.27,42.28) -- (104.27,56.78) ; \draw   (90.27,42.28) -- (104.52,42.28)(90.27,56.28) -- (104.52,56.28) ; \draw    ;
\draw  [draw opacity=0][fill={rgb, 255:red, 126; green, 211; blue, 33 }  ,fill opacity=1 ] (146.27,42.26) -- (160.52,42.26) -- (160.52,56.76) -- (146.27,56.76) -- cycle ; \draw   (146.27,42.26) -- (146.27,56.76)(160.27,42.26) -- (160.27,56.76) ; \draw   (146.27,42.26) -- (160.52,42.26)(146.27,56.26) -- (160.52,56.26) ; \draw    ;

\end{tikzpicture}
        }
        \centerline{$\mathcal{M}_i^1=\mathcal{M}_i^2$}
        \centerline{(a)}
    \end{minipage}
         \begin{minipage}[t]{0.01\linewidth}
        \centering
        \centerline{}
        \end{minipage}
    \begin{minipage}[t]{\figwidth}
        \centering
        \centerline{

\tikzset{every picture/.style={line width=0.75pt}} 

\begin{tikzpicture}[x=0.75pt,y=0.75pt,yscale=-1,xscale=1]

\draw (35.25,35.25) node  {\includegraphics[width=52.5pt,height=52.5pt]{figures/S2_img.png}};
\draw (125.25,35.25) node  {\includegraphics[width=52.5pt,height=52.5pt]{figures/S1_img.png}};
\draw  [draw opacity=0] (0.24,0.27) -- (70.73,0.23) -- (70.77,70.73) -- (0.28,70.77) -- cycle ; \draw   (0.24,0.27) -- (0.28,70.77)(14.24,0.26) -- (14.28,70.76)(28.24,0.25) -- (28.28,70.75)(42.24,0.25) -- (42.28,70.75)(56.24,0.24) -- (56.28,70.74)(70.24,0.23) -- (70.28,70.73) ; \draw   (0.24,0.27) -- (70.73,0.23)(0.25,14.27) -- (70.74,14.23)(0.25,28.27) -- (70.75,28.23)(0.26,42.27) -- (70.75,42.23)(0.27,56.27) -- (70.76,56.23)(0.28,70.27) -- (70.77,70.23) ; \draw    ;
\draw  [draw opacity=0] (90.25,0.29) -- (160.74,0.25) -- (160.78,70.75) -- (90.29,70.79) -- cycle ; \draw   (90.25,0.29) -- (90.29,70.79)(104.25,0.28) -- (104.29,70.78)(118.25,0.27) -- (118.29,70.77)(132.25,0.26) -- (132.29,70.76)(146.25,0.26) -- (146.29,70.76)(160.25,0.25) -- (160.29,70.75) ; \draw   (90.25,0.29) -- (160.74,0.25)(90.26,14.29) -- (160.75,14.25)(90.26,28.29) -- (160.76,28.25)(90.27,42.29) -- (160.76,42.25)(90.28,56.29) -- (160.77,56.25)(90.29,70.29) -- (160.78,70.25) ; \draw    ;
\draw  [draw opacity=0][fill={rgb, 255:red, 126; green, 211; blue, 33 }  ,fill opacity=1 ] (28.28,56.25) -- (42.53,56.25) -- (42.53,70.75) -- (28.28,70.75) -- cycle ; \draw   (28.28,56.25) -- (28.28,70.75)(42.28,56.25) -- (42.28,70.75) ; \draw   (28.28,56.25) -- (42.53,56.25)(28.28,70.25) -- (42.53,70.25) ; \draw    ;
\draw  [draw opacity=0][fill={rgb, 255:red, 208; green, 2; blue, 27 }  ,fill opacity=1 ] (56.27,42.23) -- (70.52,42.23) -- (70.52,56.73) -- (56.27,56.73) -- cycle ; \draw   (56.27,42.23) -- (56.27,56.73)(70.27,42.23) -- (70.27,56.73) ; \draw   (56.27,42.23) -- (70.52,42.23)(56.27,56.23) -- (70.52,56.23) ; \draw    ;
\draw  [draw opacity=0][fill={rgb, 255:red, 126; green, 211; blue, 33 }  ,fill opacity=1 ] (0.25,14.27) -- (14.5,14.27) -- (14.5,28.77) -- (0.25,28.77) -- cycle ; \draw   (0.25,14.27) -- (0.25,28.77)(14.25,14.27) -- (14.25,28.77) ; \draw   (0.25,14.27) -- (14.5,14.27)(0.25,28.27) -- (14.5,28.27) ; \draw    ;
\draw  [draw opacity=0][fill={rgb, 255:red, 126; green, 211; blue, 33 }  ,fill opacity=1 ] (42.25,0.25) -- (56.5,0.25) -- (56.5,14.75) -- (42.25,14.75) -- cycle ; \draw   (42.25,0.25) -- (42.25,14.75)(56.25,0.25) -- (56.25,14.75) ; \draw   (42.25,0.25) -- (56.5,0.25)(42.25,14.25) -- (56.5,14.25) ; \draw    ;
\draw  [draw opacity=0][fill={rgb, 255:red, 208; green, 2; blue, 27 }  ,fill opacity=1 ] (28.25,28.25) -- (42.5,28.25) -- (42.5,42.75) -- (28.25,42.75) -- cycle ; \draw   (28.25,28.25) -- (28.25,42.75)(42.25,28.25) -- (42.25,42.75) ; \draw   (28.25,28.25) -- (42.5,28.25)(28.25,42.25) -- (42.5,42.25) ; \draw    ;
\draw  [draw opacity=0][fill={rgb, 255:red, 208; green, 2; blue, 27 }  ,fill opacity=1 ] (0.26,42.27) -- (14.51,42.27) -- (14.51,56.77) -- (0.26,56.77) -- cycle ; \draw   (0.26,42.27) -- (0.26,56.77)(14.26,42.27) -- (14.26,56.77) ; \draw   (0.26,42.27) -- (14.51,42.27)(0.26,56.27) -- (14.51,56.27) ; \draw    ;
\draw  [draw opacity=0][fill={rgb, 255:red, 126; green, 211; blue, 33 }  ,fill opacity=1 ] (90.26,14.29) -- (104.51,14.29) -- (104.51,28.79) -- (90.26,28.79) -- cycle ; \draw   (90.26,14.29) -- (90.26,28.79)(104.26,14.29) -- (104.26,28.79) ; \draw   (90.26,14.29) -- (104.51,14.29)(90.26,28.29) -- (104.51,28.29) ; \draw    ;
\draw  [draw opacity=0][fill={rgb, 255:red, 208; green, 2; blue, 27 }  ,fill opacity=1 ] (104.27,28.28) -- (118.52,28.28) -- (118.52,42.78) -- (104.27,42.78) -- cycle ; \draw   (104.27,28.28) -- (104.27,42.78)(118.27,28.28) -- (118.27,42.78) ; \draw   (104.27,28.28) -- (118.52,28.28)(104.27,42.28) -- (118.52,42.28) ; \draw    ;
\draw  [draw opacity=0][fill={rgb, 255:red, 208; green, 2; blue, 27 }  ,fill opacity=1 ] (132.26,14.26) -- (146.51,14.26) -- (146.51,28.76) -- (132.26,28.76) -- cycle ; \draw   (132.26,14.26) -- (132.26,28.76)(146.26,14.26) -- (146.26,28.76) ; \draw   (132.26,14.26) -- (146.51,14.26)(132.26,28.26) -- (146.51,28.26) ; \draw    ;
\draw  [draw opacity=0][fill={rgb, 255:red, 126; green, 211; blue, 33 }  ,fill opacity=1 ] (118.29,56.26) -- (132.54,56.26) -- (132.54,70.76) -- (118.29,70.76) -- cycle ; \draw   (118.29,56.26) -- (118.29,70.76)(132.29,56.26) -- (132.29,70.76) ; \draw   (118.29,56.26) -- (132.54,56.26)(118.29,70.26) -- (132.54,70.26) ; \draw    ;
\draw  [draw opacity=0][fill={rgb, 255:red, 208; green, 2; blue, 27 }  ,fill opacity=1 ] (132.27,42.26) -- (146.52,42.26) -- (146.52,56.76) -- (132.27,56.76) -- cycle ; \draw   (132.27,42.26) -- (132.27,56.76)(146.27,42.26) -- (146.27,56.76) ; \draw   (132.27,42.26) -- (146.52,42.26)(132.27,56.26) -- (146.52,56.26) ; \draw    ;
\draw  [draw opacity=0][fill={rgb, 255:red, 126; green, 211; blue, 33 }  ,fill opacity=1 ] (132.25,0.26) -- (146.5,0.26) -- (146.5,14.76) -- (132.25,14.76) -- cycle ; \draw   (132.25,0.26) -- (132.25,14.76)(146.25,0.26) -- (146.25,14.76) ; \draw   (132.25,0.26) -- (146.5,0.26)(132.25,14.26) -- (146.5,14.26) ; \draw    ;

\end{tikzpicture}

        }
        \centerline{$|\mathcal{M}_i^1 \cap \mathcal{M}_i^2| \geq 0$}
        \centerline{(b)}
    \end{minipage}
\begin{minipage}[t]{0.01\linewidth}
        \centering
        \centerline{}
        \end{minipage}
    \begin{minipage}[t]{\figwidth}
        \centering
        \centerline{

\tikzset{every picture/.style={line width=0.75pt}} 

\begin{tikzpicture}[x=0.75pt,y=0.75pt,yscale=-1,xscale=1]

\draw (35.25,35.25) node  {\includegraphics[width=52.5pt,height=52.5pt]{figures/S2_img.png}};
\draw (125.25,35.25) node  {\includegraphics[width=52.5pt,height=52.5pt]{figures/S1_img.png}};
\draw  [draw opacity=0] (0.24,0.27) -- (70.73,0.23) -- (70.77,70.73) -- (0.28,70.77) -- cycle ; \draw   (0.24,0.27) -- (0.28,70.77)(14.24,0.26) -- (14.28,70.76)(28.24,0.25) -- (28.28,70.75)(42.24,0.25) -- (42.28,70.75)(56.24,0.24) -- (56.28,70.74)(70.24,0.23) -- (70.28,70.73) ; \draw   (0.24,0.27) -- (70.73,0.23)(0.25,14.27) -- (70.74,14.23)(0.25,28.27) -- (70.75,28.23)(0.26,42.27) -- (70.75,42.23)(0.27,56.27) -- (70.76,56.23)(0.28,70.27) -- (70.77,70.23) ; \draw    ;
\draw  [draw opacity=0] (90.25,0.29) -- (160.74,0.25) -- (160.78,70.75) -- (90.29,70.79) -- cycle ; \draw   (90.25,0.29) -- (90.29,70.79)(104.25,0.28) -- (104.29,70.78)(118.25,0.27) -- (118.29,70.77)(132.25,0.26) -- (132.29,70.76)(146.25,0.26) -- (146.29,70.76)(160.25,0.25) -- (160.29,70.75) ; \draw   (90.25,0.29) -- (160.74,0.25)(90.26,14.29) -- (160.75,14.25)(90.26,28.29) -- (160.76,28.25)(90.27,42.29) -- (160.76,42.25)(90.28,56.29) -- (160.77,56.25)(90.29,70.29) -- (160.78,70.25) ; \draw    ;
\draw  [draw opacity=0][fill={rgb, 255:red, 208; green, 2; blue, 27 }  ,fill opacity=1 ] (28.28,56.25) -- (42.53,56.25) -- (42.53,70.75) -- (28.28,70.75) -- cycle ; \draw   (28.28,56.25) -- (28.28,70.75)(42.28,56.25) -- (42.28,70.75) ; \draw   (28.28,56.25) -- (42.53,56.25)(28.28,70.25) -- (42.53,70.25) ; \draw    ;
\draw  [draw opacity=0][fill={rgb, 255:red, 208; green, 2; blue, 27 }  ,fill opacity=1 ] (56.27,42.23) -- (70.52,42.23) -- (70.52,56.73) -- (56.27,56.73) -- cycle ; \draw   (56.27,42.23) -- (56.27,56.73)(70.27,42.23) -- (70.27,56.73) ; \draw   (56.27,42.23) -- (70.52,42.23)(56.27,56.23) -- (70.52,56.23) ; \draw    ;
\draw  [draw opacity=0][fill={rgb, 255:red, 208; green, 2; blue, 27 }  ,fill opacity=1 ] (0.25,14.27) -- (14.5,14.27) -- (14.5,28.77) -- (0.25,28.77) -- cycle ; \draw   (0.25,14.27) -- (0.25,28.77)(14.25,14.27) -- (14.25,28.77) ; \draw   (0.25,14.27) -- (14.5,14.27)(0.25,28.27) -- (14.5,28.27) ; \draw    ;
\draw  [draw opacity=0][fill={rgb, 255:red, 208; green, 2; blue, 27 }  ,fill opacity=1 ] (42.25,0.25) -- (56.5,0.25) -- (56.5,14.75) -- (42.25,14.75) -- cycle ; \draw   (42.25,0.25) -- (42.25,14.75)(56.25,0.25) -- (56.25,14.75) ; \draw   (42.25,0.25) -- (56.5,0.25)(42.25,14.25) -- (56.5,14.25) ; \draw    ;
\draw  [draw opacity=0][fill={rgb, 255:red, 208; green, 2; blue, 27 }  ,fill opacity=1 ] (28.25,28.25) -- (42.5,28.25) -- (42.5,42.75) -- (28.25,42.75) -- cycle ; \draw   (28.25,28.25) -- (28.25,42.75)(42.25,28.25) -- (42.25,42.75) ; \draw   (28.25,28.25) -- (42.5,28.25)(28.25,42.25) -- (42.5,42.25) ; \draw    ;
\draw  [draw opacity=0][fill={rgb, 255:red, 208; green, 2; blue, 27 }  ,fill opacity=1 ] (0.26,42.27) -- (14.51,42.27) -- (14.51,56.77) -- (0.26,56.77) -- cycle ; \draw   (0.26,42.27) -- (0.26,56.77)(14.26,42.27) -- (14.26,56.77) ; \draw   (0.26,42.27) -- (14.51,42.27)(0.26,56.27) -- (14.51,56.27) ; \draw    ;
\draw  [draw opacity=0][fill={rgb, 255:red, 208; green, 2; blue, 27 }  ,fill opacity=1 ] (104.25,0.27) -- (118.5,0.27) -- (118.5,14.77) -- (104.25,14.77) -- cycle ; \draw   (104.25,0.27) -- (104.25,14.77)(118.25,0.27) -- (118.25,14.77) ; \draw   (104.25,0.27) -- (118.5,0.27)(104.25,14.27) -- (118.5,14.27) ; \draw    ;
\draw  [draw opacity=0][fill={rgb, 255:red, 208; green, 2; blue, 27 }  ,fill opacity=1 ] (118.27,14.27) -- (132.52,14.27) -- (132.52,28.77) -- (118.27,28.77) -- cycle ; \draw   (118.27,14.27) -- (118.27,28.77)(132.27,14.27) -- (132.27,28.77) ; \draw   (118.27,14.27) -- (132.52,14.27)(118.27,28.27) -- (132.52,28.27) ; \draw    ;
\draw  [draw opacity=0][fill={rgb, 255:red, 208; green, 2; blue, 27 }  ,fill opacity=1 ] (146.26,28.25) -- (160.51,28.25) -- (160.51,42.75) -- (146.26,42.75) -- cycle ; \draw   (146.26,28.25) -- (146.26,42.75)(160.26,28.25) -- (160.26,42.75) ; \draw   (146.26,28.25) -- (160.51,28.25)(146.26,42.25) -- (160.51,42.25) ; \draw    ;
\draw  [draw opacity=0][fill={rgb, 255:red, 208; green, 2; blue, 27 }  ,fill opacity=1 ] (132.29,42.26) -- (146.54,42.26) -- (146.54,56.76) -- (132.29,56.76) -- cycle ; \draw   (132.29,42.26) -- (132.29,56.76)(146.29,42.26) -- (146.29,56.76) ; \draw   (132.29,42.26) -- (146.54,42.26)(132.29,56.26) -- (146.54,56.26) ; \draw    ;
\draw  [draw opacity=0][fill={rgb, 255:red, 208; green, 2; blue, 27 }  ,fill opacity=1 ] (104.28,42.27) -- (118.53,42.27) -- (118.53,56.77) -- (104.28,56.77) -- cycle ; \draw   (104.28,42.27) -- (104.28,56.77)(118.28,42.27) -- (118.28,56.77) ; \draw   (104.28,42.27) -- (118.53,42.27)(104.28,56.27) -- (118.53,56.27) ; \draw    ;
\draw  [draw opacity=0][fill={rgb, 255:red, 208; green, 2; blue, 27 }  ,fill opacity=1 ] (90.25,56.25) -- (104.5,56.25) -- (104.5,70.75) -- (90.25,70.75) -- cycle ; \draw   (90.25,56.25) -- (90.25,70.75)(104.25,56.25) -- (104.25,70.75) ; \draw   (90.25,56.25) -- (104.5,56.25)(90.25,70.25) -- (104.5,70.25) ; \draw    ;

\end{tikzpicture}
        }
        \centerline{$\mathcal{M}_i^1 \cap \mathcal{M}_i^2 = \emptyset$}
        \centerline{(c)}
    \end{minipage}
    \caption{An illustration of three different multi-modal masking correspondences: (a) identical; (b) random; and (c) disjoint. For each one, if the same local areas are masked out on images from different sensors, they are shown in green. Otherwise, they are shown in red. }
    \label{fig:masking}
\end{figure*}
\subsection{Basics on Masked Autoencoders}
Let $\mathcal{P}_i^j=\{\boldsymbol{p}_{i,n}^j| \cap_n \boldsymbol{p}_{i,n}^j = \emptyset, \cup_n \boldsymbol{p}_{i,n}^j=\boldsymbol{x}_i^j\}$ be the set of all non-overlapping patches of the image $\boldsymbol{x}_i^j$, where each patch is the size of $K\times K$ pixels. MAEs aim to learn the representation of $\boldsymbol{x}_i^j$ by reconstructing its masked patches $\mathcal{P}_{i,\mathcal{M}}^j=\{\boldsymbol{p}_{i,n}^j| n \in \mathcal{M}_i^j\}$ conditioned on the remaining unmasked patches $\mathcal{P}_{i,\mathcal{U}}^j = \mathcal{P}_i^j \setminus \mathcal{P}_{i,\mathcal{M}}^j$, where $\mathcal{M}_i^j$ is the set of indices denoting the masked patches. To achieve this, an MAE is composed of a pair of encoder $e$ and decoder $d$. For both encoder and decoder, vision transformers (ViTs) are employed in MAEs. ViTs operate on the lower dimensional embeddings of image patches, which are denoted as tokens. To this end, the unmasked image patches are first linearly transformed into embeddings, which are then combined with sinusoidal positional encodings that result in token embeddings. For $\boldsymbol{x}_i^j$, the encoder of the MAE operates only on the token embeddings of unmasked image patches to generate the latent representations $\mathcal{Z}_{i,\mathcal{U}}^j = \{\boldsymbol{z}^j_{i,n}|\boldsymbol{z}^j_{i,n}=e(\boldsymbol{p}_{i,n}^j), \boldsymbol{p}_{i,n}^j\in\mathcal{P}_{i,\mathcal{U}}^j\}$ of those patches (i.e., $\mathcal{Z}_{i,\mathcal{U}}^j = e(\mathcal{P}_{i,\mathcal{U}}^j)$). The decoder of the MAE reconstructs the masked patches $\hat{\mathcal{P}}_{i,\mathcal{M}}^j=\{\hat{\boldsymbol{p}}_{i,n}^j| n \in \mathcal{M}_i^j\}$ conditioned on the latent representations of the unmasked patches and the special mask token embeddings $\boldsymbol{z}_{i,m}^j$ in place of the masked patches (i.e., $\hat{\mathcal{P}}_{i,\mathcal{M}}^j = d(\mathcal{Z}_{i,\mathcal{U}}^j, \boldsymbol{z}_{i,m}^j)$). The last layer of the MAE decoder is a linear projection, in which the number of outputs equals to the total number of pixels associated to an image patch. Accordingly, the learning objective of MAE is the uni-modal reconstruction loss $\mathcal{L}_{\text{UMR}}$ between the masked patches and the corresponding reconstructed patches. For a given image $\boldsymbol{x}_i^j$, this can be written as the mean squared error computed between $\hat{\mathcal{P}}_{i,\mathcal{M}}^j$ and $\mathcal{P}_{i,\mathcal{M}}^j$ as follows:
\begin{equation}
       \mathcal{L}_{\text{UMR}}(\boldsymbol{x}_i^j) = 
    \frac{1}{|\mathcal{M}_i^j|}\sum_{n \in \mathcal{M}_i^j} ||d(\mathcal{Z}_{i,\mathcal{U}}^j, \boldsymbol{z}_{i,m}^j) - \boldsymbol{p}_{i,n}^j||^2.
\end{equation}
Training the considered MAE with this objective leads to modeling the characteristics of RS images associated with a single modality (i.e., either $j=1$ or $j=2$ for all the considered images). Although this has been found to be effective for learning the intra-modal image characteristics on a given image modality, the vanilla MAEs can not be directly utilized for multi-modal image archives. For the characterization of RS image representations on multi-modal image archives, one could employ a different MAE specific to each image modality in separate learning procedures. However, this prevents to model inter-modal RS image characteristics, which requires accurate modeling of the shared semantic content across image modalities in the same learning procedure. We would like to highlight that not only intra-modal but also inter-modal RS image characteristics are of great importance for sensor-agnostic CBIR problems in RS~\cite{Sumbul:2022}. 

\subsection{Adaptation of MAEs for Sensor-Agnostic Image Retrieval}
To simultaneously model inter-modal and intra-modal image characteristics on multi-modal RS image archives, we adapt MAEs for sensor-agnostic CBIR problems and present cross-sensor masked autoencoders (\basemae{s}). For a given multi-modal RS image pair $(\boldsymbol{x}_i^1, \boldsymbol{x}_i^2)$, the proposed \basemae{s} aim to learn the representations of $\boldsymbol{x}_i^1$ and $\boldsymbol{x}_i^2$ based on not only uni-modal reconstruction objectives as in (1) but also cross-modal reconstruction objectives as follows. \basemae{s} learn the representation of $\boldsymbol{x}_i^1$ by reconstructing its masked patches $\mathcal{P}_{i,\mathcal{M}}^1$ conditioned on both: 1) its unmasked patches $\mathcal{P}_{i,\mathcal{U}}^1$ (i.e., uni-modal reconstruction); and 2) the unmasked patches $\mathcal{P}_{i,\mathcal{U}}^2$ of $\boldsymbol{x}_i^2$ (i.e., cross-modal reconstruction). At the same time, \basemae{s} reconstruct $\mathcal{P}_{i,\mathcal{M}}^2$ conditioned on both $\mathcal{P}_{i,\mathcal{U}}^2$ and $\mathcal{P}_{i,\mathcal{U}}^1$ for learning the representation of $\boldsymbol{x}_i^2$. To achieve this, we adapt MAEs into~\basemae{s}~by applying adaptations on: 1) image masking; 2) ViT architecture; and iii) masked image modeling. Fig.~\ref{fig:overall_model} shows an illustration of the~\basemae{s}, while our adaptations on MAEs are explained in detail in the following. 

\subsubsection{Adaptation on Image Masking}
In MAEs, each training sample $\boldsymbol{x}_i^j$ is associated with one set $\mathcal{M}_i^j$ of indices, which defines which patches are masked out. This is generally randomly generated based on a pre-defined masking ratio. However, for~\basemae{s}, each training sample $(\boldsymbol{x}_i^1, \boldsymbol{x}_i^2)$ is a pair of images, which requires to define two sets $\mathcal{M}_i^1$ and $\mathcal{M}_i^2$ of indices. Although each set can be still randomly defined based on the masking ratio, the relation between $\mathcal{M}_i^1$ and $\mathcal{M}_i^2$ (i.e., multi-modal masking correspondence) affects the hardness of the cross-modal reconstruction objective for~\basemae{s}. We define three different multi-modal masking correspondences, each of which is associated with a different hardness level for cross-modal reconstruction. If $\mathcal{M}_i^1=\mathcal{M}_i^2$ (which is denoted as identical multi-modal masking correspondence), the image patches acquired by different sensors on the same geographical area are masked out during encoding. This is associated with the hardest level of cross-modal reconstruction, where $\mathcal{P}_{i,\mathcal{U}}^j$ and $\mathcal{P}_{i,\mathcal{M}}^k$ do not share any geographical areas for $(j,k)=(1,2)$ and $(2,1)$. If $\mathcal{M}_i^1\cap \mathcal{M}_i^2 = \emptyset$ (which is denoted as disjoint multi-modal masking correspondence), the image patches acquired by different sensors on different geographical areas are masked out during encoding. This is associated with the easiest level of cross-modal reconstruction, where $\mathcal{P}_{i,\mathcal{U}}^j$ and $\mathcal{P}_{i,\mathcal{M}}^k$ are the image patches acquired by different sensors on the same geographical areas for $(j,k)=(1,2)$ and $(2,1)$. If $|\mathcal{M}_i^1\cap \mathcal{M}_i^2| \geq 0$ (which is denoted as random multi-modal masking correspondence), the image patches acquired by different sensors on the same or different geographical areas can be masked out during encoding. This is associated with the medium level of cross-modal reconstruction, where $\mathcal{P}_{i,\mathcal{U}}^j$ and $\mathcal{P}_{i,\mathcal{M}}^k$ may share the same geographical areas for $(j,k)=(1,2)$ and $(2,1)$. These different types of multi-modal masking correspondences in~\basemae~are illustrated in Fig. \ref{fig:masking}.

\begin{figure*}[t]
   \newcommand{\figwidth}{0.233\linewidth}
    \newcommand{\figheight}{1.15in}
    \centering
     \begin{minipage}[t]{\figwidth}
        \centering
        \centerline{

\tikzset{every picture/.style={line width=0.75pt}} 

\begin{tikzpicture}[x=0.75pt,y=0.75pt,yscale=-1,xscale=1]

\draw [color={rgb, 255:red, 134; green, 24; blue, 24 }  ,draw opacity=1 ][line width=1.5]    (100.5,60.25) -- (114.25,60.25) ;
\draw [shift={(118.25,60.25)}, rotate = 180] [fill={rgb, 255:red, 134; green, 24; blue, 24 }  ,fill opacity=1 ][line width=0.08]  [draw opacity=0] (8.13,-3.9) -- (0,0) -- (8.13,3.9) -- cycle    ;
\draw [color={rgb, 255:red, 134; green, 24; blue, 24 }  ,draw opacity=1 ][line width=1.5]    (40.75,60.25) -- (55.75,60.45) ;
\draw [shift={(59.75,60.5)}, rotate = 180.75] [fill={rgb, 255:red, 134; green, 24; blue, 24 }  ,fill opacity=1 ][line width=0.08]  [draw opacity=0] (8.13,-3.9) -- (0,0) -- (8.13,3.9) -- cycle    ;
\draw  [fill={rgb, 255:red, 155; green, 0; blue, 155 }  ,fill opacity=0.3 ] (0.89,35.36) -- (40.54,45) -- (40.54,76.11) -- (0.89,85.75) -- cycle ;
\draw  [fill={rgb, 255:red, 155; green, 155; blue, 155 }  ,fill opacity=0.3 ] (60.39,35.36) -- (100.04,44.75) -- (100.04,76.36) -- (60.39,85.75) -- cycle ;
\draw  [fill={rgb, 255:red, 155; green, 255; blue, 155 }  ,fill opacity=0.3 ] (158.44,84.83) -- (118.83,75.25) -- (118.95,44.95) -- (158.64,35.69) -- cycle ;
\draw [draw opacity=0]   (1,120) -- (30.25,120) ;

\draw (2.75,42.25) node [anchor=north west][inner sep=0.75pt]  [font=\scriptsize] [align=left] {Sensor\\Common\\Encoder};
\draw (62.25,42.25) node [anchor=north west][inner sep=0.75pt]  [font=\scriptsize] [align=left] {Cross\\Sensor\\Encoder};
\draw (120.25,44) node [anchor=north west][inner sep=0.75pt]  [font=\scriptsize] [align=left] {Sensor\\Common\\Decoder};

\end{tikzpicture}
        }
        \centerline{\small(a) \mbox{\basemae}-CECD}
    \end{minipage}
         \begin{minipage}[t]{0.01\linewidth}
        \centering
        \centerline{}
        \end{minipage}
    \begin{minipage}[t]{\figwidth}
        \centering
        \centerline{

\tikzset{every picture/.style={line width=0.75pt}} 

\begin{tikzpicture}[background rectangle/.style={fill=gray!10}, show background rectangle,x=0.75pt,y=0.75pt,yscale=-1,xscale=1]

\draw [color={rgb, 255:red, 134; green, 24; blue, 24 }  ,draw opacity=1 ][line width=1.5]    (106.25,65) -- (106.25,25.5) -- (114.25,25.33) ;
\draw [shift={(118.25,25.25)}, rotate = 178.81] [fill={rgb, 255:red, 134; green, 24; blue, 24 }  ,fill opacity=1 ][line width=0.08]  [draw opacity=0] (8.13,-3.9) -- (0,0) -- (8.13,3.9) -- cycle    ;
\draw [color={rgb, 255:red, 134; green, 24; blue, 24 }  ,draw opacity=1 ][line width=1.5]    (100.5,60.25) -- (106.25,60.25) ;
\draw [color={rgb, 255:red, 134; green, 24; blue, 24 }  ,draw opacity=1 ][line width=1.5]    (106.25,55.5) -- (106.25,96) -- (114,95.84) ;
\draw [shift={(118,95.75)}, rotate = 178.78] [fill={rgb, 255:red, 134; green, 24; blue, 24 }  ,fill opacity=1 ][line width=0.08]  [draw opacity=0] (8.13,-3.9) -- (0,0) -- (8.13,3.9) -- cycle    ;
\draw  [fill={rgb, 255:red, 155; green, 155; blue, 155 }  ,fill opacity=0.3 ] (60.39,35.36) -- (100.04,44.75) -- (100.04,76.36) -- (60.39,85.75) -- cycle ;
\draw  [fill={rgb, 255:red, 155; green, 55; blue, 0 }  ,fill opacity=0.3 ] (158.44,50.83) -- (118.83,41) -- (118.95,10.45) -- (158.64,0.94) -- cycle ;
\draw  [fill={rgb, 255:red, 155; green, 55; blue, 0 }  ,fill opacity=0.3 ] (158.44,120.58) -- (118.83,111) -- (118.95,80.7) -- (158.64,71.44) -- cycle ;
\draw [color={rgb, 255:red, 134; green, 24; blue, 24 }  ,draw opacity=1 ][line width=1.5]    (41.25,60) -- (56.25,60.2) ;
\draw [shift={(60.25,60.25)}, rotate = 180.75] [fill={rgb, 255:red, 134; green, 24; blue, 24 }  ,fill opacity=1 ][line width=0.08]  [draw opacity=0] (8.13,-3.9) -- (0,0) -- (8.13,3.9) -- cycle    ;
\draw  [fill={rgb, 255:red, 155; green, 0; blue, 155 }  ,fill opacity=0.3 ] (1.39,35.11) -- (41.04,44.75) -- (41.04,75.86) -- (1.39,85.5) -- cycle ;

\draw (62.25,42.25) node [anchor=north west][inner sep=0.75pt]  [font=\scriptsize] [align=left] {Cross\\Sensor\\Encoder};
\draw (120.25,10) node [anchor=north west][inner sep=0.75pt]  [font=\scriptsize] [align=left] {Sensor\\Specific\\Decoder};
\draw (120.25,80) node [anchor=north west][inner sep=0.75pt]  [font=\scriptsize] [align=left] {Sensor\\Specific\\Decoder};
\draw (3.25,42.25) node [anchor=north west][inner sep=0.75pt]  [font=\scriptsize] [align=left] {Sensor\\Common\\Encoder};

\end{tikzpicture}

        }
        \centerline{\small(b) \mbox{\basemae}-CESD}
    \end{minipage}
\begin{minipage}[t]{0.01\linewidth}
        \centering
        \centerline{}
        \end{minipage}
    \begin{minipage}[t]{\figwidth}
        \centering
        \centerline{

\tikzset{every picture/.style={line width=0.75pt}} 

\begin{tikzpicture}[x=0.75pt,y=0.75pt,yscale=-1,xscale=1]

\draw  [fill={rgb, 255:red, 0; green, 0; blue, 155 }  ,fill opacity=0.3 ] (0.6,0.61) -- (40.25,10.25) -- (40.25,41.36) -- (0.6,51) -- cycle ;
\draw [color={rgb, 255:red, 134; green, 24; blue, 24 }  ,draw opacity=1 ][line width=1.5]    (40.75,95.75) -- (46.75,95.75) -- (46.75,55.5) ;
\draw [color={rgb, 255:red, 134; green, 24; blue, 24 }  ,draw opacity=1 ][line width=1.5]    (40.75,25.5) -- (46.75,25.5) -- (46.75,65.25) ;
\draw [color={rgb, 255:red, 134; green, 24; blue, 24 }  ,draw opacity=1 ][line width=1.5]    (47.75,60.5) -- (55.75,60.5) ;
\draw [shift={(59.75,60.5)}, rotate = 180] [fill={rgb, 255:red, 134; green, 24; blue, 24 }  ,fill opacity=1 ][line width=0.08]  [draw opacity=0] (8.13,-3.9) -- (0,0) -- (8.13,3.9) -- cycle    ;
\draw  [fill={rgb, 255:red, 0; green, 0; blue, 155 }  ,fill opacity=0.3 ] (0.64,70.36) -- (40.29,80) -- (40.29,111.11) -- (0.64,120.75) -- cycle ;
\draw  [fill={rgb, 255:red, 155; green, 155; blue, 155 }  ,fill opacity=0.3 ] (60.39,35.36) -- (100.04,44.75) -- (100.04,76.36) -- (60.39,85.75) -- cycle ;
\draw [color={rgb, 255:red, 134; green, 24; blue, 24 }  ,draw opacity=1 ][line width=1.5]    (100.5,60.75) -- (114.25,60.75) ;
\draw [shift={(118.25,60.75)}, rotate = 180] [fill={rgb, 255:red, 134; green, 24; blue, 24 }  ,fill opacity=1 ][line width=0.08]  [draw opacity=0] (8.13,-3.9) -- (0,0) -- (8.13,3.9) -- cycle    ;
\draw  [fill={rgb, 255:red, 155; green, 255; blue, 155 }  ,fill opacity=0.3 ] (158.44,85.33) -- (118.83,75.75) -- (118.95,45.45) -- (158.64,36.19) -- cycle ;

\draw (2.46,7.5) node [anchor=north west][inner sep=0.75pt]  [font=\scriptsize] [align=left] {Sensor\\Specific\\Encoder};
\draw (2.5,77.25) node [anchor=north west][inner sep=0.75pt]  [font=\scriptsize] [align=left] {Sensor\\Specific\\Encoder};
\draw (62.25,42.25) node [anchor=north west][inner sep=0.75pt]  [font=\scriptsize] [align=left] {Cross\\Sensor\\Encoder};
\draw (120.25,45) node [anchor=north west][inner sep=0.75pt]  [font=\scriptsize] [align=left] {Sensor\\Common\\Decoder};

\end{tikzpicture}

        }
        \centerline{\small(c) \mbox{\basemae}-SECD}
    \end{minipage}
    \begin{minipage}[t]{0.01\linewidth}
        \centering
        \centerline{}
        \end{minipage}
    \begin{minipage}[t]{\figwidth}
        \centering
        \centerline{

\tikzset{every picture/.style={line width=0.75pt}} 

\begin{tikzpicture}[background rectangle/.style={fill=gray!10}, show background rectangle,x=0.75pt,y=0.75pt,yscale=-1,xscale=1]

\draw  [fill={rgb, 255:red, 0; green, 0; blue, 155 }  ,fill opacity=0.3 ] (0.6,0.61) -- (40.25,10.25) -- (40.25,41.36) -- (0.6,51) -- cycle ;
\draw [color={rgb, 255:red, 134; green, 24; blue, 24 }  ,draw opacity=1 ][line width=1.5]    (40.75,95.75) -- (46.75,95.75) -- (46.75,55.5) ;
\draw [color={rgb, 255:red, 134; green, 24; blue, 24 }  ,draw opacity=1 ][line width=1.5]    (40.75,25.5) -- (46.75,25.5) -- (46.75,65.25) ;
\draw [color={rgb, 255:red, 134; green, 24; blue, 24 }  ,draw opacity=1 ][line width=1.5]    (106.25,65) -- (106.75,25.25) -- (114.25,25.25) ;
\draw [shift={(118.25,25.25)}, rotate = 180] [fill={rgb, 255:red, 134; green, 24; blue, 24 }  ,fill opacity=1 ][line width=0.08]  [draw opacity=0] (8.13,-3.9) -- (0,0) -- (8.13,3.9) -- cycle    ;
\draw [color={rgb, 255:red, 134; green, 24; blue, 24 }  ,draw opacity=1 ][line width=1.5]    (100.5,60.25) -- (106.25,60.25) ;
\draw [color={rgb, 255:red, 134; green, 24; blue, 24 }  ,draw opacity=1 ][line width=1.5]    (47.75,60.5) -- (55.75,60.5) ;
\draw [shift={(59.75,60.5)}, rotate = 180] [fill={rgb, 255:red, 134; green, 24; blue, 24 }  ,fill opacity=1 ][line width=0.08]  [draw opacity=0] (8.13,-3.9) -- (0,0) -- (8.13,3.9) -- cycle    ;
\draw [color={rgb, 255:red, 134; green, 24; blue, 24 }  ,draw opacity=1 ][line width=1.5]    (106.25,55.5) -- (106.25,96) -- (114,95.84) ;
\draw [shift={(118,95.75)}, rotate = 178.78] [fill={rgb, 255:red, 134; green, 24; blue, 24 }  ,fill opacity=1 ][line width=0.08]  [draw opacity=0] (8.13,-3.9) -- (0,0) -- (8.13,3.9) -- cycle    ;
\draw  [fill={rgb, 255:red, 0; green, 0; blue, 155 }  ,fill opacity=0.3 ] (0.64,70.36) -- (40.29,80) -- (40.29,111.11) -- (0.64,120.75) -- cycle ;
\draw  [fill={rgb, 255:red, 155; green, 155; blue, 155 }  ,fill opacity=0.3 ] (60.39,35.36) -- (100.04,44.75) -- (100.04,76.36) -- (60.39,85.75) -- cycle ;
\draw  [fill={rgb, 255:red, 155; green, 55; blue, 0 }  ,fill opacity=0.3 ] (158.44,50.83) -- (118.83,41) -- (118.95,10.45) -- (158.64,0.94) -- cycle ;
\draw  [fill={rgb, 255:red, 155; green, 55; blue, 0 }  ,fill opacity=0.3 ] (158.44,120.58) -- (118.83,111) -- (118.95,80.7) -- (158.64,71.44) -- cycle ;

\draw (2.46,7.5) node [anchor=north west][inner sep=0.75pt]  [font=\scriptsize] [align=left] {Sensor\\Specific\\Encoder};
\draw (2.5,77.25) node [anchor=north west][inner sep=0.75pt]  [font=\scriptsize] [align=left] {Sensor\\Specific\\Encoder};
\draw (62.25,42.25) node [anchor=north west][inner sep=0.75pt]  [font=\scriptsize] [align=left] {Cross\\Sensor\\Encoder};
\draw (120.25,9.5) node [anchor=north west][inner sep=0.75pt]  [font=\scriptsize] [align=left] {Sensor\\Specific\\Decoder};
\draw (120.25,80) node [anchor=north west][inner sep=0.75pt]  [font=\scriptsize] [align=left] {Sensor\\Specific\\Decoder};

\end{tikzpicture}

        }
        \centerline{\small(d) \mbox{\basemae}-SESD}
    \end{minipage}
    \caption{An illustration of four different \mbox{\basemae} models. A \mbox{\basemae} is composed of a multi-sensor encoder, a cross-sensor encoder and a multi-sensor decoder, each of which is based on ViTs. For (a) \mbox{\basemae}-CECD and (b) \mbox{\basemae}-CESD, the multi-sensor encoder employs a sensor-common encoder for producing the latent representations of the unmasked patches. For (c) \mbox{\basemae}-SECD and (d) \mbox{\basemae}-SESD, the multi-sensor encoder employs sensor-specific encoders, where ViT encoders with different parameters are utilized for different image modalities. For (a) \mbox{\basemae}-CECD and (c) \mbox{\basemae}-SECD, the multi-sensor decoder employs a sensor-common decoder for reconstruction. For (b) \mbox{\basemae}-CESD and (d) \mbox{\basemae}-SESD, the multi-sensor decoder employs sensor-specific decoders, where ViT decoders with different parameters are utilized for different image modalities.}
    \label{fig:arch}
\end{figure*}
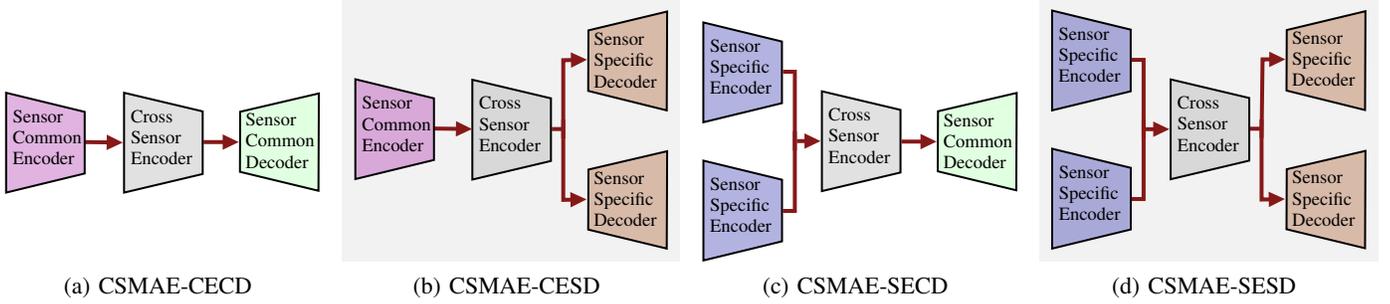
\subsubsection{Adaptation on ViT Architecture}
A \basemae~is composed of a multi-sensor encoder, a cross-sensor encoder and a multi-sensor decoder, each of which is based on ViTs. To obtain token embeddings for the considered ViTs, a different linear transformation is utilized for each image modality, while using the same sinusoidal positional encoding.

The multi-sensor encoder of the \basemae~may employ a sensor-common encoder, where the same ViT generates the latent representations of the unmasked patches of both image modalities. Alternatively, the multi-sensor encoder may employ sensor-specific encoders, where ViT encoders with different parameters are utilized for different image modalities. Once a sensor-common encoder is used to obtain the latent representations of each modality, there can be a risk of inaccurately modeling sensor-specific image characteristics. To prevent this, sensor-specific encoders can be used in the multi-sensor encoder of the \basemae~at the cost of an increase in the number of model parameters. The cross-sensor encoder of \basemae~takes the output of the multi-sensor encoder and maps the latent representations of different modalities into a common embedding space (i.e., $\mathcal{Z}_{i,\mathcal{U}}^1$ and $\mathcal{Z}_{i,\mathcal{U}}^2$ for $\boldsymbol{x}_i^1$ and $\boldsymbol{x}_i^2$, respectively). This is particularly important for sensor-agnostic CBIR problems, where the image representations need to be comparable to each other even across different image modalities. 

For each image of $(\boldsymbol{x}_i^1, \boldsymbol{x}_i^2)$, the multi-sensor decoder $d$ of the \mbox{\basemae} takes the output of the cross-sensor encoder and reconstructs the corresponding masked patches two times. Initially, the reconstruction is conditioned on the unmasked patches of the image itself, as in the vanilla MAE. Then, the patches are also reconstructed based on the unmasked patches from the other image modality (i.e., $\hat{\mathcal{P}}_{i,\mathcal{M}}^j = d(\mathcal{Z}_{i,\mathcal{U}}^k, \boldsymbol{z}_{i,m}^k) \quad \forall j \neq k$). The multi-sensor decoder of \basemae may employ a sensor-common decoder, where the same ViT decodes the latent representations of both $\boldsymbol{x}_i^1$ and $\boldsymbol{x}_i^2$ for reconstruction. Alternatively, the multi-sensor decoder of \mbox{\basemae} may employ sensor-specific decoders, where ViT decoders with different parameters are utilized for different image modalities. For both configurations, reconstruction in image pixel space is achieved by different linear projection layers for different image modalities. The selection of sensor-common decoder may limit to accurately reconstruct information specific to a sensor. This can be hindered by using sensor-specific decoders in the multi-sensor decoder at the cost of an increase in the number of model parameters. Based on these alternatives, we define four~\basemae~models: 1)~\basemae~with sensor common encoder and decoder (\basemae-CECD); 2)~\basemae~with sensor common encoder and sensor specific decoders (\basemae-CESD); 3)~\basemae~with sensor specific encoders and sensor common decoder (\basemae-SECD); and 4)~\basemae~with sensor specific encoders and decoders (\basemae-SESD). These models are shown in Fig. \ref{fig:arch}. 

\subsubsection{Adaptation on Masked Image Modeling}
Once the reconstructed image patches $\hat{\mathcal{P}}_{i,\mathcal{M}}^1$ and $\hat{\mathcal{P}}_{i,\mathcal{M}}^2$ of $(\boldsymbol{x}_i^1, \boldsymbol{x}_i^2)$ are obtained, the corresponding learning objective \mbox{$\mathcal{L}_{\text{\basemae}}$} of \mbox{\basemae{s}} can be written based on not only $\mathcal{L}_{\text{UMR}}$, but also the cross-modal reconstruction loss $\mathcal{L}_{\text{CMR}}$ as follows:
\begin{equation}
\begin{aligned}
    \mathcal{L}_{\text{CMR}}(\boldsymbol{x}_i^j) &= 
    \frac{1}{|\mathcal{M}_i^j|}\sum_{n \in \mathcal{M}_i^j} \!\!\!||d(\mathcal{Z}_{i,\mathcal{U}}^k, \boldsymbol{z}_{i,m}^k) - \boldsymbol{p}_{i,n}^j||^2 \\
    \mathcal{L}_{\text{\basemae}}(\boldsymbol{x}_i^1, \boldsymbol{x}_i^2) &= \sum_{j\in\{1,2\}}\mathcal{L}_{\text{UMR}}(\boldsymbol{x}_i^j) + \mathcal{L}_{\text{CMR}}(\boldsymbol{x}_i^j),
\end{aligned}
\end{equation}
where $k=2$ if $j=1$ and vice versa. In this way, the characterization of RS image representations is achieved by learning to reconstruct both: i) images within the same image modality; and ii) images across different image modalities. This leads to modeling not only intra-modal but also inter-modal RS image characteristics.

We would like to note that the cross-modal reconstruction objective in (2) allows to implicitly model image similarities across different image modalities. However, this may be limited for accurately encoding inter-modal image similarities on the cross-sensor encoder if the considered image modalities are significantly different from each other. To prevent this and to explicitly model inter-modal image similarities, by following~\cite{Sumbul:2022}, inter-modal discrepancy elimination loss function $\mathcal{L}_{\text{MDE}}$ can be included in the overall training loss in addition to (2). For a given set $\mathcal{B}$ of training sample indices, $\mathcal{L}_{\text{MDE}}$ is written as follows:
\begin{equation}
    \mathcal{L}_{\text{MDE}}(\mathcal{B}) = -\frac{1}{|\mathcal{B}|} \sum_{i\in \mathcal{B}} \text{log}(1+e^{S(\boldsymbol{c}_{i}^1, \boldsymbol{c}_{i}^2)}),
\end{equation}
where $S(\cdot, \cdot)$ measures cosine similarity, $\boldsymbol{c}_i^j$ denotes the overall latent representation of $\boldsymbol{x}_i^j$ (i.e., feature vector). $\boldsymbol{c}_i^j$ can be obtained by either: i) applying global average pooling (GAP) over $\mathcal{Z}_{i,\mathcal{U}}^j$; or ii) using the special [CLS] token. $\mathcal{L}_{\text{MDE}}$ enforces to have a small angular distance between the latent representations of images acquired by different sensors on the same geographical area. 

For latent similarity preservation, as an alternative to $\mathcal{L}_{\text{MDE}}$, the mutual information maximization loss function $\mathcal{L}_{\text{MIM}}$ from~\cite{Sumbul:2022} can be also utilized. $\mathcal{L}_{\text{MIM}}$ is defined based on the normalized temperature-scaled cross entropy~\cite{Bachman:2019} as follows:
\begin{equation}
\begin{aligned}
\ell^i(j,k) &= -\text{log}(\frac{e^{S(\boldsymbol{c}_i^j, \boldsymbol{c}_i^k)/\tau}}{\sum_{q \in \mathcal{B}}\mathds{1}_{[q \neq i]}e^{S(\boldsymbol{c}_i^j, \boldsymbol{c}_q^k)/\tau}}),\\
\mathcal{L}_{\text{MIM}}(\mathcal{B}) &= \frac{1}{2|\mathcal{B}|}\sum_{i\in\mathcal{B}} \ell^i(1,2) + \ell^i(2,1),\\
\end{aligned}
\end{equation}
where $\mathds{1}$ is the indicator function and $\tau$ is the temperature parameter. It is worth noting that $\mathcal{L}_{\text{MDE}}$ considers latent similarity preservation on one training sample independently of the other training samples. However, $\mathcal{L}_{\text{MIM}}$ enforces to maximize the latent similarity of each training sample compared to the other training samples (i.e., $|\mathcal{B}| -1$ negative samples in the mini-batch)~\cite{Poole:2019}.

After training the considered~\basemae~model by minimizing either (2) or (2), (3) and (4), the latent representations $\{\boldsymbol{c}_i^1\}$ and $\{\boldsymbol{c}_i^2\}$ for $\forall\boldsymbol{x}_i^1\in\mathcal{X}^1$ and $\forall\boldsymbol{x}_i^2\in\mathcal{X}^2$ are obtained. Then, for a given query image, the $k$ most similar images within each image modality or across different image modalities are selected by comparing the corresponding image representations based on the $k$-nn algorithm.
\begin{table*}[t]
\renewcommand{\arraystretch}{0.01}
\setlength\tabcolsep{2pt}
\caption{The Hyperparameter Selection Range and The Variations of the~\basemae~Models Considered for the Sensitivity Analysis and the Ablation Study of the Experimental Results.}
\label{tab:cmmae_variations}
\centering
\scriptsize
\begin{tabular}{@{}lccccccccc@{}}
\hline
\thead[c]{\scriptsize{}Model\\\scriptsize{}Name} & \thead[c]{\scriptsize{}$K$ for\\\scriptsize{}Patch Size} & \thead[c]{\scriptsize{}Masking\\\scriptsize{}Ratio} & \thead[c]{\scriptsize{}ViT Variant for\\\scriptsize{}Multi-Sensor\\\scriptsize{}Encoder} & \thead[c]{\scriptsize{}Cross-sensor\\\scriptsize{}Encoder\\\scriptsize{}Depth} & \thead[c]{\scriptsize{}Feature\\\scriptsize{} Vector\\\scriptsize{}Type} & $\tau$ & \thead[c]{\scriptsize{}Multi-Modal\\\scriptsize{}Masking\\\scriptsize{}Correspondence} & \thead[c]{\scriptsize{}Reconstruction\\\scriptsize{}Loss} & \thead[c]{\scriptsize{}Inter-Modal\\\scriptsize{}Latent Similarity\\\scriptsize{}Preservation} \\ \hline
\basemae-CECD & \thead[c]{\scriptsize{}\{8,10,12,15,\\\scriptsize{}20,24,30,60\}} & \{10\%$,\dots,$90\%\} & \thead[c]{\scriptsize{}\{ViT-Ti12,ViT-S12,\\\scriptsize{}ViT-B12,ViT-L24\}} & 2 & \{[CLS], GAP\} & \{0.1,0.5,1,5,10\} & \thead[c]{\scriptsize{}\{Identical, Random,\\\scriptsize{}Disjoint\}} & \thead[c]{\scriptsize{}\{$\mathcal{L}_{\text{UMR}}$, $\mathcal{L}_{\text{CMR}}$,\\\scriptsize{}$\mathcal{L}_{\text{UMR}}$ \& $\mathcal{L}_{\text{CMR}}$\}} & \thead[c]{\scriptsize{}\{\xmark, $\mathcal{L}_{\text{MDE}}$, $\mathcal{L}_{\text{MIM}}$,\\\scriptsize{}$\mathcal{L}_{\text{MDE}}$ \& $\mathcal{L}_{\text{MIM}}$\}}  \\
\basemae-CESD & 15 & 50\% & ViT-B12 & 2& \{[CLS], GAP\} & 0.5 & \thead[c]{\scriptsize{}\{Identical, Random,\\\scriptsize{}Disjoint\}} & \thead[c]{\scriptsize{}\{$\mathcal{L}_{\text{UMR}}$, $\mathcal{L}_{\text{CMR}}$,\\\scriptsize{}$\mathcal{L}_{\text{UMR}}$ \& $\mathcal{L}_{\text{CMR}}$\}} & \thead[c]{\scriptsize{}\{\xmark, $\mathcal{L}_{\text{MDE}}$, $\mathcal{L}_{\text{MIM}}$,\\\scriptsize{}$\mathcal{L}_{\text{MDE}}$ \& $\mathcal{L}_{\text{MIM}}$\}} \\
\basemae-SECD & 15 & 50\% & ViT-B12 & \{2,4,6,8,10\}& \{[CLS], GAP\} & 0.5 & \thead[c]{\scriptsize{}\{Identical, Random,\\\scriptsize{}Disjoint\}} & \thead[c]{\scriptsize{}\{$\mathcal{L}_{\text{UMR}}$, $\mathcal{L}_{\text{CMR}}$,\\\scriptsize{}$\mathcal{L}_{\text{UMR}}$ \& $\mathcal{L}_{\text{CMR}}$\}} & \thead[c]{\scriptsize{}\{\xmark, $\mathcal{L}_{\text{MDE}}$, $\mathcal{L}_{\text{MIM}}$,\\\scriptsize{}$\mathcal{L}_{\text{MDE}}$ \& $\mathcal{L}_{\text{MIM}}$\}} \\
\basemae-SESD & 15 & 50\% & ViT-B12 &2& \{[CLS], GAP\} & 0.5 &  \thead[c]{\scriptsize{}\{Identical, Random,\\\scriptsize{}Disjoint\}} & \thead[c]{\scriptsize{}\{$\mathcal{L}_{\text{UMR}}$, $\mathcal{L}_{\text{CMR}}$,\\\scriptsize{}$\mathcal{L}_{\text{UMR}}$ \& $\mathcal{L}_{\text{CMR}}$\}} & \thead[c]{\scriptsize{}\{\xmark, $\mathcal{L}_{\text{MDE}}$, $\mathcal{L}_{\text{MIM}}$,\\\scriptsize{}$\mathcal{L}_{\text{MDE}}$ \& $\mathcal{L}_{\text{MIM}}$\}} \\

\hline
\end{tabular}
\end{table*}
\section{Data Set Description and Experimental Setup}
\label{sec:exp_setup}
\subsection{Data Set Description}
We conducted experiments on the BigEarthNet benchmark archive \cite{BigEarthNetMM} that includes 590,326 multi-modal image pairs. Each pair in BigEarthNet includes one Sentinel-1 (denoted as S1) SAR image and one Sentinel-2 (denoted as S2) multispectral image acquired on the same geographical area. Each pair is associated with one or more class labels (i.e., multi-labels) based on the 19 class nomenclature. For the experiments, we separately stacked: i) the VV and VH bands of S1 images; and ii) the S2 bands associated with 10m and 20m spatial resolution, while bicubic interpolation was applied to 20m bands. 

To perform experiments, we selected two sets of image pairs. For the first set, we selected the 270,470 BigEarthNet image pairs acquired over 10 European countries during summer and autumn (which is denoted as BEN-270K). For the second set, we selected the 14,832 BigEarthNet image pairs acquired over Serbia during summer (which is denoted as BEN-14K). Then, we divided each set into training (52\%), validation (24\%) and test (24\%) subsets. The training subsets of both sets were used to perform training. The validation subset of BEN-14K was used to select query images, while images were retrieved from the test subset of BEN-14K.

\subsection{Experimental Setup}
For all the~\basemae~models, we utilized different ViT variants for the multi-sensor and the cross-sensor encoders that are shown in Table~\ref{tab:cmmae_variations}. When a sensor-common encoder is utilized (i.e.,~\basemae-CECD or~\basemae-CESD), the overall architecture of the multi-sensor encoder becomes identical to the architecture of the considered ViT variant. When sensor-specific encoders are utilized (i.e.,~\basemae-SECD or~\basemae-SESD), the depth of specific encoders defines how many hidden layers of the considered ViT variant are replicated two times. For the multi-sensor decoder, we employed either one or two ViTs depending on the use of sensor-specific decoders. For both cases, by following~\cite{He:2022}, we considered the ViT with 8 transformer blocks and 16 attention heads that operates with the embedding dimension of 512. We trained all the models for 150 epochs with a mini-batch size of 128. For training, the AdamW optimizer was utilized with the initial learning rate of 10\textsuperscript{-4}, which follows the linear warmup schedule with cosine annealing. All the experiments were conducted on NVIDIA A100 GPUs. 

We carried out different kinds of experiments to: 1) perform a sensitivity analysis with respect to different hyperparameters; 2) conduct an ablation study of the introduced~\basemae~models; and 3) compare the~\basemae~models with other approaches. For the sensitivity analysis and the ablation study, different values of several hyperparameters for different~\basemae~models were considered. All the considered variations are shown in Table~\ref{tab:cmmae_variations}. For the comparison with other approaches, we consider two scenarios: i) training different approaches from scratch on the same training set with the \mbox{\basemae} models; and ii) pretrained models on natural images, while always using the same ViT variant with our models. For the first scenario, we compare the \mbox{\basemae} models with: 1) the vanilla masked autoencoder (MAE) \mbox{\cite{He:2022}}; 2) the MAE with the rotated varied-size attention (denoted as MAE-RVSA) \mbox{\cite{Wang:2023}}; 3) the self-supervised cross-modal image retrieval method (denoted as SS-CMIR) \mbox{\cite{Sumbul:2022}}; and 4) the masked vision language modeling method (denoted as MaskVLM) \mbox{\cite{Kwon:2023}}. We would like to note that MAE and MAE-RVSA can not be directly trained for multi-modal RS image archives, and thus can not be utilized for cross-sensor CBIR. For such approaches, we trained one model specific to each image modality that results in MAE (S1), MAE (S2), MAE-RVSA (S1) and MAE-RVSA (S2). Then, for cross-sensor CBIR, we obtained the image features associated with each image modality from the corresponding modality-specific models. For a fair comparison, we used the same number of training epochs in MAE, MAE-RVSA, SS-CMIR and MaskVLM with our \mbox{\basemae} models. For the second scenario, we compare the \mbox{\basemae} models with: 1) MAE \mbox{\cite{He:2022}} pretrained on ImageNet-1K (IN1K) dataset \mbox{\cite{Deng:2009}}; 2) a simple framework for masked image modeling (SimMIM) \mbox{\cite{Xie:2022}} pretrained on IN1K dataset; 3) Supervised ViT \mbox{\cite{Dosovitskiy:2021}} pretrained on ImageNet-21K (IN21K) dataset \mbox{\cite{Ridnik:2021:2}}; and 4) CLIP \mbox{\cite{Radford:2021}} pretrained on Yahoo Flickr Creative Commons 100 Million (YFCC100M) dataset \mbox{\cite{Bart:2016}}.

In the experiments, the results of \textit{S1} $\rightarrow$ \textit{S1}, \textit{S2} $\rightarrow$ \textit{S2}, \textit{S1} $\rightarrow$ \textit{S2} and \textit{S2} $\rightarrow$ \textit{S1} CBIR tasks are given in terms of $F_1$ score (which is a widely used metric for CBIR tasks when the considered images are associated with multi-labels as in BigEarthNet \mbox{\cite{Sumbul:2022, BigEarthNet-S2, Sumbul:2022:2, Sumbul:2021, Sumbul:2021:2}}). $F_1$ score is the harmonic mean of precision and recall. Thus, it allows to represent both the precision and recall of the CBIR results with a single metric. This prevents to increase the complexity of the experimental results with multiple metrics for four different CBIR tasks. For a detailed description of the considered metric, the reader is referred to \mbox{\cite{Chaudhuri:2018}}. For each CBIR task, the first image modality (e.g., \textit{S1} for the \textit{S1} $\rightarrow$ \textit{S2} task) denotes the associated modality of a query image, while the second one (e.g., \textit{S2} for the \textit{S1} $\rightarrow$ \textit{S2} task) denotes the image modality associated with retrieved images. The retrieval performance was assessed on top-10 retrieved images.

\begin{table}[t]
\renewcommand{\arraystretch}{1.1}
\setlength\tabcolsep{9pt}
\caption{$F_1$ Scores (\%) Obtained by \basemae-CECD~(ViT-B12/K) When Different Patch Sizes K$\times$K are Used Under Random Multi-Modal Masking Correspondence (BEN-14K).}
\label{tab:sensitivity_patch_size}
\centering
\begin{tabular}{@{}cccccc@{}}
\hline
\multirow{2}{*}{\thead[c]{Patch\\Size}} & \multicolumn{2}{c}{\thead[c]{Uni-Modal CBIR}}& & \multicolumn{2}{c}{\thead[c]{Cross-Modal CBIR}} \\ \cline{2-3} \cline{5-6}  
& \textit{S1} $\rightarrow$ \textit{S1} & \textit{S2} $\rightarrow$ \textit{S2} & & \textit{S1} $\rightarrow$ \textit{S2} & \textit{S2} $\rightarrow$ \textit{S1} \\ \hline
8$\times$8 & \textbf{70.43} & \textbf{74.34} & & 44.08 & 43.06 \\ 
10$\times$10 & 70.07 & 74.13 & & 52.36 & 56.77 \\ 
12$\times$12 & 69.91 & 73.82 & & 45.47 & 52.18\\ 
15$\times$15 & 69.89 & 74.21 & & \textbf{64.27} & 52.91 \\ 
20$\times$20 & 68.09 & 74.30 & & 57.48 & 58.60\\ 
24$\times$24 & 67.63 & 74.24 & & 53.57 & \textbf{59.84} \\ 
30$\times$30 & 66.28 & 73.69 & & 50.49 & 56.31\\ 
60$\times$60 & 58.02 & 70.45 & & 46.44 & 54.45\\ \hline
\end{tabular}
\end{table}
\section{Experimental Results}
\label{sec:result}
\subsection{Sensitivity Analysis}
In this sub-section, we present the results of the sensitivity analysis for the~\basemae~models in terms of different patch sizes, masking ratios, ViT variants for multi-sensor encoders, depths of cross-sensor encoders and feature vector types used for inter-modal similarity preservation. We would like to note that, for the sensitivity analysis, we present the results of the~\basemae-CECD and~\basemae-SECD models trained on BEN-14K due to space constraints since we observed the similar behaviors of all the models trained on both BEN-14K and BEN-270K. Unless it is stated differently, for the sensitivity analysis, the considered models are exploited without inter-modal latent similarity preservation, while random multi-modal masking correspondence is used.  

\subsubsection{Patch Size} In the first set of trials, we analyzed the effect of the patch size used in~\basemae{s}. Table \ref{tab:sensitivity_patch_size} shows the results of the~\basemae-CECD model when ViT-B12 is employed for the multi-sensor encoder. One can observe from the table that~\basemae-CECD achieves the highest $F_1$ scores for uni-modal CBIR tasks when the image patch size is set to 8$\times$8. However, this significantly reduces the performance of~\basemae-CECD on cross-modal CBIR tasks, for which~\basemae-CECD provides the higher scores when bigger image patches (e.g., 15$\times$15 for \textit{S1} $\rightarrow$ \textit{S2} retrieval task) are utilized. A small patch size results in smaller adjacent and unmasked regions in images. By this way, the \mbox{\basemae}-CECD model focuses more on uni-modal reconstruction objective than cross-modal reconstruction objective. This results in sub-optimal convergence for cross-modal reconstruction objective, and thus lower $F_1$ scores for cross-modal CBIR tasks. This is due to the relative complexity of cross-modal reconstruction objective compared to uni-modal reconstruction objective. These results show that there is not a single value of the patch size, for which~\basemae-CECD leads to the highest accuracies on both uni-modal and cross-modal CBIR tasks. It is noted that~\basemae-CECD achieves the best performance in terms of average $F_1$ score of all retrieval tasks when the image patch size is set to 15$\times$15. Accordingly, for the rest of the experiments, we utilized the image patches with the size of 15$\times$15 in the~\basemae models.

\begin{table}[t]
\renewcommand{\arraystretch}{1.1}
\setlength\tabcolsep{9pt}
\caption{$F_1$ Scores (\%) Obtained by the \basemae-CECD~(ViT-B12/15) When Different Masking Ratios are Used Under Random Multi-Modal Masking Correspondence (BEN-14K).}
\label{tab:sensitivity_masking_ratio}
\centering
\begin{tabular}{@{}cccccc@{}}
\hline
\multirow{2}{*}{\thead[c]{Masking\\Ratio}} & \multicolumn{2}{c}{\thead[c]{Uni-Modal CBIR}}& & \multicolumn{2}{c}{\thead[c]{Cross-Modal CBIR}}\\ \cline{2-3} \cline{5-6}   
& \textit{S1} $\rightarrow$ \textit{S1} & \textit{S2} $\rightarrow$ \textit{S2} & & \textit{S1} $\rightarrow$ \textit{S2} & \textit{S2} $\rightarrow$ \textit{S1} \\ \hline
10\% & 67.64 & 73.93 & & 52.41 & 57.88 \\ 
20\% & 68.45 & 74.05 & & 54.54 & 57.59 \\ 
30\% & 68.97 & 74.02 & & 56.25 & 54.73 \\ 
40\% & 69.55 & 74.12 & & 60.83 & 54.09 \\ 
50\% & \textbf{69.89} & \textbf{74.21} & & \textbf{64.27} & 52.91 \\ 
60\% & 69.79 & 74.03 & & 60.93 & 59.12 \\ 
70\% & 69.55 & 74.10 & & 61.43 & 59.10 \\ 
80\% & 69.19 & 74.31 & & 62.00 & 62.11 \\ 
90\% & 67.97 & 73.86 & & 58.02 & \textbf{64.34} \\ \hline
\end{tabular}
\end{table}
\subsubsection{Masking Ratio} In the second set of trials, we assessed the effect of the masking ratio used in~\basemae{s}. Table \ref{tab:sensitivity_masking_ratio} shows the results of the~\basemae-CECD model when ViT-B12 is used for the multi-sensor encoder. One can see from the table that when the considered masking ratio is set to 50\%, \basemae-CECD achieves the highest $F_1$ scores on three out of the four CBIR tasks. On the \textit{S2} $\rightarrow$ \textit{S1} retrieval task,~\basemae-CECD leads to the highest scores when the masking ratio is set to 90\%, for which the performance of~\basemae-CECD significantly decreases for the other CBIR tasks. Accordingly, the masking ratio of 50\% is exploited in the~\basemae~models for the rest of the experiments.

\begin{table}[t]
\renewcommand{\arraystretch}{1.1}
\setlength\tabcolsep{4pt}
\caption{$F_1$ Scores (\%) Obtained by \basemae-CECD together with the Number of Parameters (NP) When Different ViT Variants Are Used for Multi-Sensor Encoder Under Random Multi-Modal Masking Correspondence (BEN-14K).}
\label{tab:sensitivity_ViT}
\centering
\begin{tabular}{@{}lcccccc@{}}
\hline
\multirow{2}{*}{\thead[c]{ViT Variant}} & \multicolumn{2}{c}{\thead[c]{Uni-Modal CBIR}} & & \multicolumn{2}{c}{\thead[c]{Cross-Modal CBIR}} & \multirow{2}{*}{\thead[c]{NP (M)}}\\ \cline{2-3} \cline{5-6}  
& \textit{S1} $\rightarrow$ \textit{S1} & \textit{S2} $\rightarrow$ \textit{S2} & & \textit{S1} $\rightarrow$ \textit{S2} & \textit{S2} $\rightarrow$ \textit{S1} & \\ \hline
ViT-Ti12/15~\cite{Touvron:2021} & 66.59 & 73.66 & & 36.68 & 49.78 & 32.57\\ 
ViT-S12/15~\cite{Touvron:2021} & 68.01 & 74.09 & & 48.60 & 49.68 & 49.14\\ 
ViT-B12/15~\cite{Dosovitskiy:2021:ViT} & \textbf{69.89} & \textbf{74.21} & & \textbf{64.27} & 52.91 & 114.15 \\ 
ViT-L24/15~\cite{Dosovitskiy:2021:ViT} & 69.35 & 73.75 & & 60.77 & \textbf{59.64} & 181.08 \\ \hline
\end{tabular}
\end{table}
\begin{table*}[t]
\renewcommand{\arraystretch}{1.1}
\setlength\tabcolsep{9pt}
\caption{$F_1$ Scores (\%) Obtained by \basemae-SECD~(ViT-B12/15) together with the Number of Parameters When Different Depths of Cross-Sensor Encoder are Used Under Random Multi-Modal Masking Correspondence (BEN-14K).}
\label{tab:sensitivity_separate_encoder}
\centering
\begin{tabular}{@{}cccccccc@{}}
\hline
\multirow{2}{*}{\thead[c]{Sensor-Specific\\Encoder Depth}} & \multirow{2}{*}{\thead[c]{Cross-Sensor\\Encoder Depth}} &  \multicolumn{2}{c}{\thead[c]{Uni-Modal CBIR}}& & \multicolumn{2}{c}{\thead[c]{Cross-Modal CBIR}}& \multirow{2}{*}{\thead[c]{Number of\\Parameters (M)}} \\ \cline{3-4} \cline{6-7}  
& & \textit{S1} $\rightarrow$ \textit{S1} & \textit{S2} $\rightarrow$ \textit{S2} & & \textit{S1} $\rightarrow$ \textit{S2} & \textit{S2} $\rightarrow$ \textit{S1} & \\ \hline
10 & 2 & \textbf{70.55} & \textbf{75.12} & & \textbf{61.67} & \textbf{66.95} & 185.03\\ 
8 & 4 & 70.49 & 75.04 & & 61.58 & 65.67 & 170.85 \\ 
6 & 6 & 70.43 & 74.80 & & 61.25 & 64.97 & 156.68 \\ 
4 & 8 & 69.73 & 74.25 & & 59.56 & 61.69 & 142.50\\ 
2 & 10 & 69.41 & 74.18 & & 57.46 & 58.72 & 128.33 \\ 
\hline
\end{tabular}
\end{table*}
\begin{table*}[t]
\renewcommand{\arraystretch}{1.1}
\setlength\tabcolsep{5pt}
\caption{$F_1$ Scores (\%) Obtained by \basemae-CECD~(ViT-B12/15) When Inter-Modal Latent Similarity Preservation is Included, Different Values of Temperature $\tau$ Are Used for $\mathcal{L}_{\text{MIM}}$ and Different Feature Vector Types [CLS] and GAP Are Used for $\mathcal{L}_{\text{MIM}}$ and $\mathcal{L}_{\text{MDE}}$ Under Random Multi-Modal Masking Correspondence (BEN-14K).}
\label{tab:sensitivity_temperature}
\centering
\begin{tabular}{@{}ccccccccccccc@{}}
\hline
\multirow{5}{*}{\thead[c]{Inter-Modal\\Latent Similarity\\Preservation}} & \multirow{5}{*}{\thead[c]{$\tau$}} & \multicolumn{11}{c}{\thead[c]{Feature Vector Type}} \\ \cline{3-13}
& & \multicolumn{5}{c}{\thead[c]{[CLS]}} & & \multicolumn{5}{c}{\thead[c]{GAP}} \\ \cline{3-7} \cline{9-13}
& & \multicolumn{2}{c}{\thead[c]{Uni-Modal CBIR}}& & \multicolumn{2}{c}{\thead[c]{Cross-Modal CBIR}} & & \multicolumn{2}{c}{\thead[c]{Uni-Modal CBIR}}& &\multicolumn{2}{c}{\thead[c]{Cross-Modal CBIR}}\\ \cline{3-4} \cline{6-7} \cline{9-10} \cline{12-13}  
& & \textit{S1} $\rightarrow$ \textit{S1} & \textit{S2} $\rightarrow$ \textit{S2} & & \textit{S1} $\rightarrow$ \textit{S2} & \textit{S2} $\rightarrow$ \textit{S1} & &\textit{S1} $\rightarrow$ \textit{S1} & \textit{S2} $\rightarrow$ \textit{S2} & & \textit{S1} $\rightarrow$ \textit{S2} & \textit{S2} $\rightarrow$ \textit{S1}\\ \hline
\multirow{5}{*}{\thead[c]{$\mathcal{L}_{\text{MIM}}$}} & 0.1 & 68.39 & 70.62 & & 69.53 & 69.25 & & 68.39 & 70.53 & & 69.30 & 69.20\\ 
 & 0.5 & \textbf{69.73} & \textbf{71.54} & & \textbf{70.08} & \textbf{70.65} & & \textbf{70.18} & \textbf{72.30} & & \textbf{70.75} & \textbf{71.39} \\ 
 & 1 & 69.06 & 70.86 & & 69.74 & 70.06 & & 69.55 & 71.40 & & 69.91 & 70.44\\ 
 & 5 & 66.40 & 68.88 & & 67.00 & 67.35 & & 67.38 & 69.74 & & 67.94 & 67.86 \\ 
 & 10 & 66.56 & 69.18 & & 67.16 & 67.12 & & 67.67 & 70.11 & & 68.07 & 68.20\\ \hline
 $\mathcal{L}_{\text{MDE}}$ & N/A & 65.59 & 73.81 & & 57.87 & 63.15 & & 69.64 & 74.02 & & 68.17 & 69.18\\ 
 \hline
\end{tabular}
\end{table*}
\subsubsection{ViT Architecture of Multi-Sensor Encoder} In the third set of trials, we analyzed the effect of the ViT variant considered in the multi-sensor encoder of~\basemae{s}. Table \ref{tab:sensitivity_ViT} shows the results of the~\basemae-CECD model. One can observe from the table that \basemae-CECD achieves the highest $F_1$ scores on all the retrieval tasks except \textit{S2} $\rightarrow$ \textit{S1} when ViT-B12 is selected as the multi-sensor encoder of the~\basemae-CECD model. In greater details, increasing the number of heads and the considered embedding dimension by using ViT-B12 instead of ViT-Ti12 or ViT-S12 leads to higher score of~\basemae-CECD on each retrieval task. However, this also leads to an increase in the number of model parameters, and thus the computational complexity of the~\basemae-CECD training. For example, in the multi-sensor encoder, using ViT-B12 leads to more than 27\% higher $F_1$ score on the \textit{S1} $\rightarrow$ \textit{S2} retrieval task compared to using ViT-Ti12 at the cost of 81.58M higher number of model parameters. However, further increasing the number of heads, embedding dimension and the depth of the multi-sensor encoder by using ViT-L24 only increases the performance of \textit{S2} $\rightarrow$ \textit{S1} retrieval task at the cost of almost 70M higher number of model parameters. Accordingly, the ViT-B12 variant is utilized in the multi-sensor encoder of the~\basemae~models for the rest of the experiments.

\subsubsection{Cross-Sensor Encoder Depth} In the fourth set of trials, we assessed the effect of cross-sensor encoder depth utilized in~\basemae{s}. Table \ref{tab:sensitivity_separate_encoder} shows the results of the~\basemae-SECD model. One can see from the table that decreasing the cross-sensor encoder depth (i.e., increasing the depth of specific encoders) in~\basemae-SECD leads to higher $F_1$ scores on all the retrieval tasks at the cost of an increase in the number of model parameters. This is more evident for cross-modal CBIR tasks. As an example, when the cross-sensor depth is set to 2,~\basemae-SECD achieves more than 8\% higher $F_1$ score on the \textit{S2} $\rightarrow$ \textit{S1} CBIR task compared to the use of 10 ViT layers in the cross-sensor encoder. To this end, for the rest of the experiments, we set the depths of the cross-sensor encoder and sensor-specific encoders to 2 and 10, respectively, for the~\basemae models.

\begin{table}[t]
\renewcommand{\arraystretch}{1.1}
\setlength\tabcolsep{2pt}
\caption{$F_1$ Scores (\%) Obtained by the \basemae-CECD, \basemae-CESD, \basemae-SECD, \basemae-SESD When Different Multi-Modal Masking Correspondences Are Used (BEN-14K).}
\label{tab:ablation_masking}
\centering
\begin{tabular}{@{}ccccccc@{}}
\hline
\multirow{4}{*}{\thead[c]{Model\\Name}} & \multirow{4}{*}{\thead[c]{Multi-Modal\\Masking\\Correspondence}} &  &&&&\\
& & \multicolumn{2}{c}{\thead[c]{Uni-Modal CBIR}} & & \multicolumn{2}{c}{\thead[c]{Cross-Modal CBIR}} \\\cline{3-4} \cline{6-7}  
& & \textit{S1} $\rightarrow$ \textit{S1} & \textit{S2} $\rightarrow$ \textit{S2} & & \textit{S1} $\rightarrow$ \textit{S2} & \textit{S2} $\rightarrow$ \textit{S1} \\ \hline
\multirow{3}{*}{\thead[c]{\basemae-CECD}}  & Identical & 69.43 & 74.08 & & 62.73 & \textbf{56.24} \\ 
& Random & \textbf{69.89} & \textbf{74.21} & & \textbf{64.27} & 52.91\\ 
& Disjoint & 69.61 & 39.10 & & 40.09 & 28.07 \\ \hline
\multirow{3}{*}{\thead[c]{\basemae-CESD}}  & Identical & 68.93 & 74.27 & & \textbf{56.48} & 53.66 \\ 
& Random & 68.77 & \textbf{74.32} & & 55.75 & \textbf{58.11}\\ 
& Disjoint & \textbf{69.19} & 39.09 & & 39.37 & 21.87 \\ \hline
\multirow{3}{*}{\thead[c]{\basemae-SECD}}  & Identical & \textbf{70.88} & \textbf{75.13} & & 58.13 & 62.51 \\ 
& Random & 70.55 & 75.12 & & \textbf{61.67} & \textbf{66.95}\\ 
& Disjoint & 70.75 & 39.27 & & 39.18 & 32.28 \\ \hline
\multirow{3}{*}{\thead[c]{\basemae-SESD}}  & Identical & \textbf{70.74} & 74.93 & & \textbf{60.01} & 62.68 \\ 
& Random & 70.49 & \textbf{75.07} & & 57.60 & \textbf{63.29}\\ 
& Disjoint & 70.62 & 39.01 & & 38.74 & 38.42 \\ 
\hline
\end{tabular}
\end{table}
\begin{table*}[t]
\renewcommand{\arraystretch}{1.1}
\setlength\tabcolsep{5pt}
\caption{$F_1$ Scores (\%) Obtained by \basemae-CECD, \basemae-CESD, \basemae-SECD, \basemae-SESD When Different Reconstruction Objectives are Used Under Random Multi-Modal Masking Correspondence Without Inter-Modal Latent Similarity Preservation (BEN-14K).}
\label{tab:ablation_reconstruction_loss}
\centering
\begin{tabular}{@{}cccccccccc@{}}
\hline
\multirow{3}{*}{\thead[c]{Model\\Name}} & & \multicolumn{2}{c}{\multirow{2}{*}{\thead[c]{Reconstruction Objective}}} & & \multicolumn{2}{c}{\multirow{2}{*}{\thead[c]{Uni-Modal CBIR}}} & & \multicolumn{2}{c}{\multirow{2}{*}{\thead[c]{Cross-Modal CBIR}}}\\
& & & & & & & & & \\  \cline{3-4} \cline{6-7} \cline{9-10}
& & Uni-Modal & Cross-Modal & & \textit{S1} $\rightarrow$ \textit{S1} & \textit{S2} $\rightarrow$ \textit{S2} & & \textit{S1} $\rightarrow$ \textit{S2} & \textit{S2} $\rightarrow$ \textit{S1}\\ \hline
\multirow{3}{*}{\basemae-CECD} & & \cmark & \xmark & & 61.58 & 71.65 & & 36.54 & 31.42 \\ 
& & \xmark & \cmark & & 69.73 & \textbf{74.39} & & 46.29 & 43.15 \\ 
& & \cmark & \cmark & & \textbf{69.89} & 74.21 & & \textbf{64.27} & \textbf{52.91} \\ \hline
\multirow{3}{*}{\basemae-CESD} & & \cmark & \xmark & & 60.74 & 70.96 & & 36.03 & 27.01 \\ 
& & \xmark & \cmark & & \textbf{69.07} & \textbf{74.50} & & 49.06 & 42.47 \\ 
& & \cmark & \cmark & & 68.77 & 74.32 & & \textbf{55.75} & \textbf{58.11} \\\hline
\multirow{3}{*}{\basemae-SECD} & & \cmark & \xmark & & 61.75 & 72.22 & & 38.13 & 24.54 \\ 
& & \xmark & \cmark & & \textbf{70.73} & 74.83 & & 53.95 & 48.40 \\ 
& & \cmark & \cmark & & 70.55 & \textbf{75.12} & &\textbf{61.67} & \textbf{66.95} \\\hline
\multirow{3}{*}{\basemae-SESD} & & \cmark & \xmark & & 60.59 & 72.06 & & 39.18 & 20.88 \\ 
& & \xmark & \cmark & & 70.31 & 75.03 & & 29.21 & 39.71 \\ 
& & \cmark & \cmark & & \textbf{70.49} & \textbf{75.07} & & \textbf{57.60} & \textbf{63.29} \\
\hline
\end{tabular}
\end{table*}

\subsubsection{Inter-Modal Latent Similarity Preservation} In the fifth set of trials, we analyzed the effect of feature vector type used in inter-modal latent similarity preservation of~\basemae{s}. Table \ref{tab:sensitivity_temperature} shows the results of the~\basemae-CECD model with inter-modal latent similarity preservation by using different feature vector types for $\mathcal{L}_{\text{MIM}}$ and $\mathcal{L}_{\text{MDE}}$ in addition to different temperature values for $\mathcal{L}_{\text{MIM}}$. One can observe from the table that inter-modal latent similarity preservation with either $\mathcal{L}_{\text{MIM}}$ or $\mathcal{L}_{\text{MDE}}$ on the global average pooling (GAP) of the patch representations of each image leads to higher $F_1$ scores than that on the [CLS] token of each image. As an example, on the \textit{S1} $\rightarrow$ \textit{S1} CBIR task,~\basemae-CECD with inter-modal latent similarity preservation through $\mathcal{L}_{\text{MDE}}$ achieves more than 4\% higher $F_1$ score when GAP is applied for the image feature vector instead of using the [CLS] token. To this end, GAP is utilized for the inter-modal latent similarity preservation of the~\basemae models for the rest of the experiments. One can also observe from Table \ref{tab:sensitivity_temperature} that when $\mathcal{L}_{\text{MIM}}$ is employed for inter-modal latent similarity preservation, using the temperature $\tau$ value of 0.5 leads to the highest results compared to the other values $\tau$ independently of the utilized feature vector type. Accordingly, for the rest of the experiments, we set $\tau$ to 0.5 for $\mathcal{L}_{\text{MIM}}$.

\subsection{Ablation Study}
In this subsection, we present the results of the ablation study of the~\basemae~models to
analyze the effectiveness of: i) different multi-modal masking correspondences; ii) reconstruction objectives $\mathcal{L}_{\text{UMR}}$ and $\mathcal{L}_{\text{CMR}}$; and iii) inter-modal latent similarity preservation. For the ablation study, we provide the results of all the~\basemae~models trained on BEN-14K. The results of the models trained on BEN-270K were inline with our conclusions from those for BEN-14K.

\begin{table*}[t]
\renewcommand{\arraystretch}{1.1}
\setlength\tabcolsep{5pt}
\caption{$F_1$ Scores (\%) Obtained by \basemae-CECD, \basemae-CESD, \basemae-SECD, \basemae-SESD When Inter-Modal Latent Similarity Preservation is Achieved with Different Loss Functions Under Random Multi-Modal Masking Correspondence (BEN-14K).}
\label{tab:ablation_intermodal}
\centering
\begin{tabular}{@{}cccccccccc@{}}
\hline
\multirow{3}{*}{\thead[c]{Model\\Name}} & & \multicolumn{2}{c}{\multirow{2}{*}{\thead[c]{Inter-Modal Latent\\Similarity Preservation}}} & & \multicolumn{2}{c}{\multirow{2}{*}{\thead[c]{Uni-Modal CBIR}}} & & \multicolumn{2}{c}{\multirow{2}{*}{\thead[c]{Cross-Modal CBIR}}}\\
& & & & & & & & & \\  \cline{3-4} \cline{6-7} \cline{9-10}
& & $\mathcal{L}_{\text{MDE}}$ & $\mathcal{L}_{\text{MIM}}$ & & \textit{S1} $\rightarrow$ \textit{S1} & \textit{S2} $\rightarrow$ \textit{S2} & & \textit{S1} $\rightarrow$ \textit{S2} & \textit{S2} $\rightarrow$ \textit{S1}\\ \hline
\multirow{4}{*}{\basemae-CECD} & & \xmark & \xmark & &  69.89 & \textbf{74.21} & & 64.27 & 52.91 \\ 
& & \xmark & \cmark & & \textbf{70.18} & 72.30 & & \textbf{70.75} & \textbf{71.39} \\ 
& & \cmark & \xmark & & 69.64 & 74.02 & & 68.17 & 69.18 \\ 
& & \cmark & \cmark & & 69.77 & 71.96 & & 70.34 & 71.06\\ \hline
\multirow{4}{*}{\basemae-CESD} & & \xmark & \xmark & & 68.77 & \textbf{74.32} & & 55.75 & 58.11 \\ 
& & \xmark & \cmark & & \textbf{70.27} & 72.51 & & \textbf{70.82} & \textbf{71.48} \\ 
& & \cmark & \xmark & & 69.53 & 74.26 & & 68.87 & 69.41 \\ 
& & \cmark & \cmark & & 69.92 & 72.16 & & 70.42 & 71.15 \\\hline
\multirow{4}{*}{\basemae-SECD} & & \xmark & \xmark & & \textbf{70.55} & \textbf{75.12} & & 61.67 & 66.95 \\ 
& & \xmark & \cmark & & 70.32 & 72.56 & & \textbf{71.06} & \textbf{71.21}\\ 
& & \cmark & \xmark & & 70.18 & 74.32 & & 68.90 & 69.71 \\ 
& & \cmark & \cmark & & 70.14 & 72.26 & & 70.83 & 71.01\\\hline
\multirow{4}{*}{\basemae-SESD} & & \xmark & \xmark & & \textbf{70.49} & \textbf{75.07} & & 57.60 & 63.29 \\ 
& & \xmark & \cmark & & 70.41 & 72.67 & & \textbf{71.16} & \textbf{71.37}\\ 
& & \cmark & \xmark & & 70.47 & 74.60 & & 69.60 & 69.49 \\ 
& & \cmark & \cmark & & 70.15 & 72.23 & & 70.70 & 71.00\\
\hline
\end{tabular}
\end{table*}

\subsubsection{Multi-Modal Masking Correspondence} In the first set of trials, we assessed the effectiveness of different multi-modal masking correspondences used in all the~\basemae~models when inter-modal latent similarity preservation is not utilized. Table \ref{tab:ablation_masking} shows the corresponding results. By observing the table, one can see that almost all the highest scores of the~\basemae~models are obtained when either identical or random multi-modal masking correspondence is considered. However, when disjoint multi-modal masking correspondence is utilized, the performance of the~\basemae~models is significantly reduced on all the CBIR tasks except \textit{S1} $\rightarrow$ \textit{S1}. This is due to the fact that disjoint multi-modal masking correspondence is associated with the easiest level of cross-modal reconstruction compared to identical and random correspondences. Such a training objective does not provide enough learning signals to model complex semantic content of multi-modal RS images, and thus leads to modeling the reconstruction of only one image modality, which is \textit{S1} in this case. In greater details, one can also observe from the table that when compared to identical multi-modal masking correspondence, random correspondence allows to achieve slightly higher $F_1$ scores on average for the~\basemae~models. Accordingly, for the rest of the experiments, random multi-modal masking correspondence is utilized with each~\basemae~model.  

\begin{table*}[!t]
\renewcommand{\arraystretch}{1.1}
\setlength\tabcolsep{5pt}
\caption{$F_1$ Scores (\%) Obtained by MAE, MAE-RVSA, SS-CMIR, MaskVLM,~\basemae-CECD,~\basemae-CESD,~\basemae-SECD and~\basemae-SESD Together with the Number of Parameters When the Training Sets of BEN-14K and BEN-270K are Considered.}
\label{tab:comparison}
\centering
\begin{tabular}{@{}ccccccccccccc@{}}
\hline
\multirow{5}{*}{\thead[c]{Model\\Name}} & \multicolumn{11}{c}{\thead[c]{Training Set}} & \multirow{5}{*}{\thead[c]{Number of\\Parameters\\(M)}}\\\cline{2-12}
& \multicolumn{5}{c}{\thead[c]{BEN-270K}} & & \multicolumn{5}{c}{\thead[c]{BEN-14K}} & \\\cline{2-6} \cline{8-12}
& \multicolumn{2}{c}{\thead[c]{Uni-Modal CBIR}}& & \multicolumn{2}{c}{\thead[c]{Cross-Modal CBIR}} & & \multicolumn{2}{c}{\thead[c]{Uni-Modal CBIR}}& &\multicolumn{2}{c}{\thead[c]{Cross-Modal CBIR}}\\ \cline{2-3} \cline{5-6} \cline{8-9} \cline{11-12}  
& \textit{S1} $\rightarrow$ \textit{S1} & \textit{S2} $\rightarrow$ \textit{S2} & & \textit{S1} $\rightarrow$ \textit{S2} & \textit{S2} $\rightarrow$ \textit{S1} & &\textit{S1} $\rightarrow$ \textit{S1} & \textit{S2} $\rightarrow$ \textit{S2} & & \textit{S1} $\rightarrow$ \textit{S2} & \textit{S2} $\rightarrow$ \textit{S1}\\ \hline
MAE (S1)~\cite{He:2022} & 70.17 & N/A & & \multirow{2}{*}{41.78} & \multirow{2}{*}{46.12} & & 60.81 & N/A & & \multirow{2}{*}{33.97} & \multirow{2}{*}{30.21}  & \multirow{2}{*}{224.87} \\ 
MAE (S2)~\cite{He:2022} & N/A & \textbf{73.97} & & & & & N/A & 72.04 & & &  & \\ 
MAE-RVSA (S1)~\cite{Wang:2023} & 52.26 & N/A & & \multirow{2}{*}{36.66} & \multirow{2}{*}{38.05} & & 55.40 & N/A & & \multirow{2}{*}{44.35} & \multirow{2}{*}{40.69}  & \multirow{2}{*}{227.75}\\ 
MAE-RVSA (S2)~\cite{Wang:2023} & N/A & 72.73 & & & & & N/A & 71.47 & & & & \\ 
SS-CMIR~\cite{Sumbul:2022} & 69.39 & 71.03 & & 69.51 & 70.52 & & 68.07 & 70.54 & & 68.86 & 69.47  & 259.07 \\ 
MaskVLM~\cite{Kwon:2023} & 69.25 & 70.85 & & 69.59 & 70.13 & & 68.10 & 71.02 & & 69.29 & 69.59 & 235.82\\ 
\basemae-CECD & 71.22 & 72.98 & & 71.61 & 72.18 & & 70.18 & 72.30 & & 70.75 & 71.39 & 114.15\\ 
\basemae-CESD & 71.00 & 72.88 & & 71.46 & 72.09 & & 70.27 & 72.51 & & 70.82 & \textbf{71.48} & 139.76\\ 
\basemae-SECD & 71.24 & 72.92 & & \textbf{71.86} & 71.94 & & 70.32 & 72.56 & & 71.06 & 71.21 & 185.03\\ 
\basemae-SESD & \textbf{71.38} & 73.16 & & 71.77 & \textbf{72.20} & & \textbf{70.41} & \textbf{72.67} & & \textbf{71.16} & 71.37 & 210.64\\ 
\hline
\end{tabular}
\end{table*}

\subsubsection{Reconstruction Objective} In the second set of trials, we analyzed the effectiveness of different reconstruction objectives considered for all the~\basemae~models when inter-modal latent similarity preservation is not utilized. Table \ref{tab:ablation_reconstruction_loss} shows the results of the~\basemae~models while using: i) only uni-modal reconstruction objective $\mathcal{L}_{\text{UMR}}$; ii) only cross-modal reconstruction objective $\mathcal{L}_{\text{CMR}}$; and iii) both $\mathcal{L}_{\text{UMR}}$ and $\mathcal{L}_{\text{CMR}}$. By assessing the table, one can observe that the joint use of uni-modal and cross-modal reconstruction objectives leads to the highest scores of the~\basemae~models on most of the CBIR tasks. This is more evident for cross-modal CBIR tasks. As an example, jointly utilizing $\mathcal{L}_{\text{UMR}}$ and $\mathcal{L}_{\text{CMR}}$ in~\basemae-SESD achieves more than 42\% higher $F_1$ score on \textit{S2} $\rightarrow$ \textit{S1} task compared to using only $\mathcal{L}_{\text{UMR}}$. These results show the effectiveness of simultaneously utilizing uni-modal and cross-modal reconstruction objectives in the~\basemae~models for sensor-agnostic retrieval problems. In greater details, one can also observe from the table that when compared to utilizing only $\mathcal{L}_{\text{UMR}}$, the single use of $\mathcal{L}_{\text{CMR}}$ increases the performance of each~\basemae~model on all the CBIR tasks. As an example,~\basemae-CESD with cross-modal reconstruction objective provides more than 8\% and 13\% higher $F_1$ scores on the CBIR tasks of \textit{S1} $\rightarrow$ \textit{S1} and \textit{S1} $\rightarrow$ \textit{S2}, respectively. This shows the effectiveness of cross-modal RS image reconstruction of~\basemae{s} for not only cross-modal CBIR, but also uni-modal CBIR.  

\begin{figure*}[!t]
    \newcommand{\figwidth}{0.11\linewidth}
    \newcommand{\figheight}{0.75in}
    \renewcommand{\fboxsep}{0pt}
    \centering
     \begin{minipage}[t]{\figwidth}
        \centering
        \centerline{\includegraphics[height=\figheight]{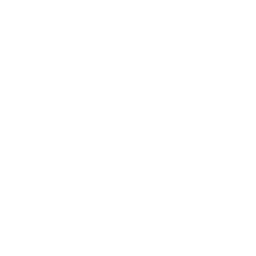}}
    \end{minipage}
     \begin{minipage}[t]{\figwidth}
        \centering
        \centerline{1{st}}\medskip
        \centerline{\includegraphics[height=\figheight]{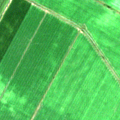}} 
    \end{minipage}
          \begin{minipage}[t]{\figwidth}
        \centering
        \centerline{2{nd}}\medskip
        \centerline{\includegraphics[height=\figheight]{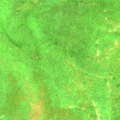}}
    \end{minipage}    
          \begin{minipage}[t]{\figwidth}
        \centering
        \centerline{3{rd}}\medskip
        \centerline{\includegraphics[height=\figheight]{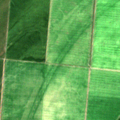}} 
    \end{minipage}
          \begin{minipage}[t]{\figwidth}
        \centering
        \centerline{4{th}}\medskip
        \centerline{\includegraphics[height=\figheight]{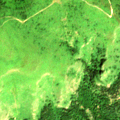}} 
        \vspace{-0.05in}\centerline{(c)}\medskip\vspace{-0.05in}
    \end{minipage}
          \begin{minipage}[t]{\figwidth}
        \centering
        \centerline{5{th}}\medskip
        \centerline{\includegraphics[height=\figheight]{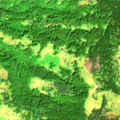}}
    \end{minipage}
\begin{minipage}[t]{\figwidth}
        \centering
        \centerline{8{th}}\medskip
        \centerline{\includegraphics[height=\figheight]{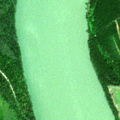}}
    \end{minipage}
     \begin{minipage}[t]{\figwidth}
        \centering
        \centerline{10{th}}\medskip
        \centerline{\includegraphics[height=\figheight]{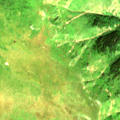}} 
    \end{minipage}
     \begin{minipage}[t]{\figwidth}
        \centering
        \centerline{\includegraphics[height=\figheight]{figures/empty.jpg}}
    \end{minipage}
          \begin{minipage}[t]{\figwidth}
        \centering
        \centerline{\includegraphics[height=\figheight]{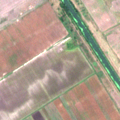}}
    \end{minipage}    
          \begin{minipage}[t]{\figwidth}
        \centering
        \centerline{\includegraphics[height=\figheight]{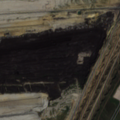}} 
    \end{minipage}
          \begin{minipage}[t]{\figwidth}
        \centering
        \centerline{\includegraphics[height=\figheight]{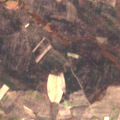}}
    \end{minipage}
          \begin{minipage}[t]{\figwidth}
        \centering
        \centerline{\includegraphics[height=\figheight]{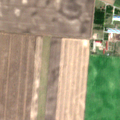}}
        \vspace{-0.05in}\centerline{(d)}\medskip\vspace{-0.05in}
    \end{minipage}
          \begin{minipage}[t]{\figwidth}
        \centering
        \centerline{\includegraphics[height=\figheight]{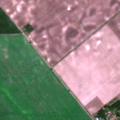}} 
    \end{minipage}
          \begin{minipage}[t]{\figwidth}
        \centering
        \centerline{\includegraphics[height=\figheight]{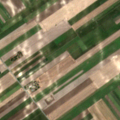}}
    \end{minipage} 
          \begin{minipage}[t]{\figwidth}
        \centering
        \centerline{\includegraphics[height=\figheight]{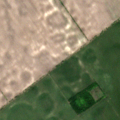}} 
    \end{minipage}
     \begin{minipage}[t]{\figwidth}
        \centering
        \centerline{\includegraphics[height=\figheight]{figures/empty.jpg}}
        
    \end{minipage}
          \begin{minipage}[t]{\figwidth}
        \centering
        \centerline{\includegraphics[height=\figheight]{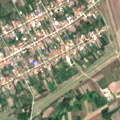}}
    \end{minipage}    
          \begin{minipage}[t]{\figwidth}
        \centering
        \centerline{\includegraphics[height=\figheight]{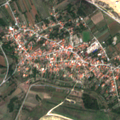}} 
    \end{minipage}
          \begin{minipage}[t]{\figwidth}
        \centering
        \centerline{\includegraphics[height=\figheight]{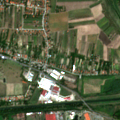}} 
    \end{minipage}
          \begin{minipage}[t]{\figwidth}
        \centering
        \centerline{\includegraphics[height=\figheight]{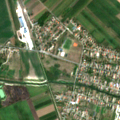}} 
        \vspace{-0.05in}\centerline{(e)}\medskip\vspace{-0.05in}
    \end{minipage}
          \begin{minipage}[t]{\figwidth}
        \centering
        \centerline{\includegraphics[height=\figheight]{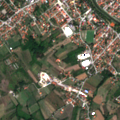}} 
    \end{minipage}
          \begin{minipage}[t]{\figwidth}
        \centering
        \centerline{\includegraphics[height=\figheight]{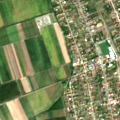}}
    \end{minipage} 
          \begin{minipage}[t]{\figwidth}
        \centering
        \centerline{\includegraphics[height=\figheight]{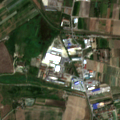}}
    \end{minipage}
     \begin{minipage}[t]{\figwidth}
        \centering
        \centerline{\includegraphics[height=\figheight]{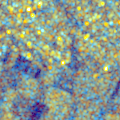}}
        \vspace{-0.05in}\centerline{(a)}\medskip\vspace{-0.05in}
    \end{minipage}
          \begin{minipage}[t]{\figwidth}
        \centering
        \centerline{\includegraphics[height=\figheight]{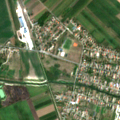}}
    \end{minipage}    
          \begin{minipage}[t]{\figwidth}
        \centering
        \centerline{\includegraphics[height=\figheight]{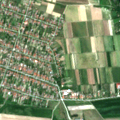}} 
    \end{minipage}
          \begin{minipage}[t]{\figwidth}
        \centering
        \centerline{\includegraphics[height=\figheight]{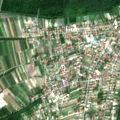}}
    \end{minipage}
          \begin{minipage}[t]{\figwidth}
        \centering
        \centerline{\includegraphics[height=\figheight]{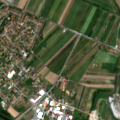}} 
        \vspace{-0.05in}\centerline{(f)}\medskip\vspace{-0.05in}
    \end{minipage}
          \begin{minipage}[t]{\figwidth}
        \centering
        \centerline{\includegraphics[height=\figheight]{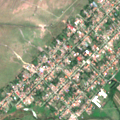}}
    \end{minipage}
          \begin{minipage}[t]{\figwidth}
        \centering
        \centerline{\includegraphics[height=\figheight]{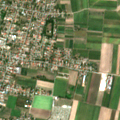}} 
    \end{minipage} 
          \begin{minipage}[t]{\figwidth}
        \centering
        \centerline{\includegraphics[height=\figheight]{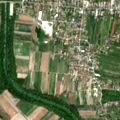}} %
    \end{minipage}
     \begin{minipage}[t]{\figwidth}
        \centering
        \centerline{\includegraphics[height=\figheight]{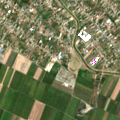}}
        \vspace{-0.05in}\centerline{(b)}\medskip\vspace{-0.05in}
    \end{minipage}
          \begin{minipage}[t]{\figwidth}
        \centering
        \centerline{\includegraphics[height=\figheight]{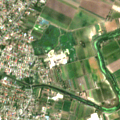}}
    \end{minipage}    
          \begin{minipage}[t]{\figwidth}
        \centering
        \centerline{\includegraphics[height=\figheight]{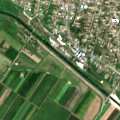}} 
    \end{minipage}
          \begin{minipage}[t]{\figwidth}
        \centering
        \centerline{\includegraphics[height=\figheight]{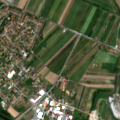}} 
    \end{minipage}
          \begin{minipage}[t]{\figwidth}
        \centering
        \centerline{\includegraphics[height=\figheight]{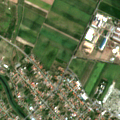}} 
        \vspace{-0.05in}\centerline{(g)}\medskip\vspace{-0.05in}
    \end{minipage}
          \begin{minipage}[t]{\figwidth}
        \centering
        \centerline{\includegraphics[height=\figheight]{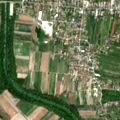}}
    \end{minipage}
          \begin{minipage}[t]{\figwidth}
        \centering
        \centerline{\includegraphics[height=\figheight]{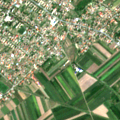}} 
    \end{minipage} 
          \begin{minipage}[t]{\figwidth}
        \centering
        \centerline{\includegraphics[height=\figheight]{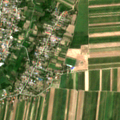}} 
    \end{minipage}
     \begin{minipage}[t]{\figwidth}
        \centering
        \centerline{\includegraphics[height=\figheight]{figures/empty.jpg}}
    \end{minipage}
          \begin{minipage}[t]{\figwidth}
        \centering
        \centerline{\includegraphics[height=\figheight]{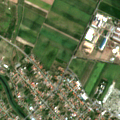}} 
    \end{minipage}    
          \begin{minipage}[t]{\figwidth}
        \centering
        \centerline{\includegraphics[height=\figheight]{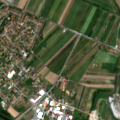}} 
    \end{minipage}
          \begin{minipage}[t]{\figwidth}
        \centering
        \centerline{\includegraphics[height=\figheight]{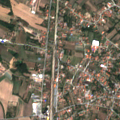}}
    \end{minipage}
          \begin{minipage}[t]{\figwidth}
        \centering
        \centerline{\includegraphics[height=\figheight]{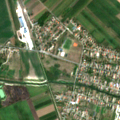}} 
        \vspace{-0.05in}\centerline{(h)}\medskip\vspace{-0.05in}
    \end{minipage}
          \begin{minipage}[t]{\figwidth}
        \centering
        \centerline{\includegraphics[height=\figheight]{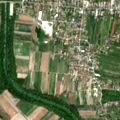}} 
    \end{minipage}
          \begin{minipage}[t]{\figwidth}
        \centering
        \centerline{\includegraphics[height=\figheight]{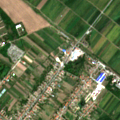}} 
    \end{minipage} 
          \begin{minipage}[t]{\figwidth}
        \centering
        \centerline{\includegraphics[height=\figheight]{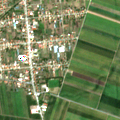}} 
    \end{minipage}
     \begin{minipage}[t]{\figwidth}
        \centering
        \centerline{\includegraphics[height=\figheight]{figures/empty.jpg}}
    \end{minipage}
          \begin{minipage}[t]{\figwidth}
        \centering
        \centerline{\includegraphics[height=\figheight]{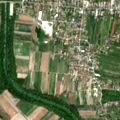}} 
    \end{minipage}    
          \begin{minipage}[t]{\figwidth}
        \centering
        \centerline{\includegraphics[height=\figheight]{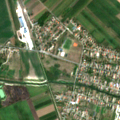}} 
    \end{minipage}
          \begin{minipage}[t]{\figwidth}
        \centering
        \centerline{\includegraphics[height=\figheight]{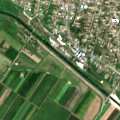}}
    \end{minipage}
          \begin{minipage}[t]{\figwidth}
        \centering
        \centerline{\includegraphics[height=\figheight]{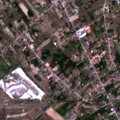}} 
        \vspace{-0.05in}\centerline{(i)}\medskip\vspace{-0.05in}
    \end{minipage}
          \begin{minipage}[t]{\figwidth}
        \centering
        \centerline{\includegraphics[height=\figheight]{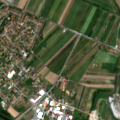}} 
    \end{minipage}
          \begin{minipage}[t]{\figwidth}
        \centering
        \centerline{\includegraphics[height=\figheight]{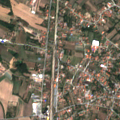}} 
    \end{minipage} 
          \begin{minipage}[t]{\figwidth}
        \centering
        \centerline{\includegraphics[height=\figheight]{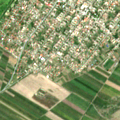}} 
    \end{minipage}
     \begin{minipage}[t]{\figwidth}
        \centering
        \centerline{\includegraphics[height=\figheight]{figures/empty.jpg}}
    \end{minipage}
          \begin{minipage}[t]{\figwidth}
        \centering
        \centerline{\includegraphics[height=\figheight]{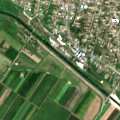}} 
    \end{minipage}    
          \begin{minipage}[t]{\figwidth}
        \centering
        \centerline{\includegraphics[height=\figheight]{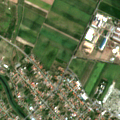}} 
    \end{minipage}
          \begin{minipage}[t]{\figwidth}
        \centering
        \centerline{\includegraphics[height=\figheight]{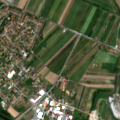}}
    \end{minipage}
          \begin{minipage}[t]{\figwidth}
        \centering
        \centerline{\includegraphics[height=\figheight]{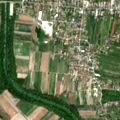}} 
        \vspace{-0.05in}\centerline{(j)}\medskip\vspace{-0.05in}
    \end{minipage}
          \begin{minipage}[t]{\figwidth}
        \centering
        \centerline{\includegraphics[height=\figheight]{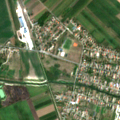}} 
    \end{minipage}
          \begin{minipage}[t]{\figwidth}
        \centering
        \centerline{\includegraphics[height=\figheight]{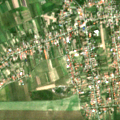}} 
    \end{minipage} 
          \begin{minipage}[t]{\figwidth}
        \centering
        \centerline{\includegraphics[height=\figheight]{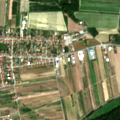}} 
    \end{minipage}
    \caption{\textit{S1} $\rightarrow$ \textit{S2} retrieval results for (a) \textit{S1} query image, (b) \textit{S2} image acquired on the same geographical area with the query image and \textit{S2} images retrieved by using: (c) MAE; (d) MAE-RVSA; (e) SS-CMIR; (f) MaskVLM; (g)~\basemae-CECD; (h)~\basemae-CESD; (i)~\basemae-SECD; and (j)~\basemae-SESD, which are trained on BEN-270K.}
    \label{fig:vis_res}
\end{figure*}
\subsubsection{Inter-Modal Latent Similarity Preservation} In the third set of trials, we analyzed the effectiveness of inter-modal latent similarity preservation employed for~\basemae{s}. Table \ref{tab:ablation_intermodal} shows the corresponding results of all the~\basemae~models, while inter-modal latent similarity preservation is achieved through: i) $\mathcal{L}_{\text{MIM}}$; ii) $\mathcal{L}_{\text{MDE}}$; and iii) both $\mathcal{L}_{\text{MIM}}$ and $\mathcal{L}_{\text{MDE}}$. By analyzing the table, one can observe that when inter-modal latent similarity preservation is utilized with either $\mathcal{L}_{\text{MIM}}$ or $\mathcal{L}_{\text{MDE}}$, each of the~\basemae-models achieves significantly higher $F_1$ scores on cross-modal CBIR tasks and similar scores on uni-modal CBIR tasks compared to not using any of $\mathcal{L}_{\text{MIM}}$ or $\mathcal{L}_{\text{MDE}}$. As an example,~\basemae-SESD with inter-modal latent similarity preservation by using $\mathcal{L}_{\text{MIM}}$ achieves almost the same performance on \textit{S1} $\rightarrow$ \textit{S1} task and more than 13\% higher $F_1$ score on \textit{S1} $\rightarrow$ \textit{S2} task compared to~\basemae-SESD without inter-modal latent similarity preservation. These show the effectiveness of inter-modal latent similarity preservation used in~\basemae{s} for cross-modal retrieval problems. In greater details, when only $\mathcal{L}_{\text{MIM}}$ is employed for inter-modal latent similarity preservation, each of the~\basemae~models provides higher $F_1$ scores on almost all the CBIR tasks compared to using only $\mathcal{L}_{\text{MDE}}$ or using both $\mathcal{L}_{\text{MIM}}$ and $\mathcal{L}_{\text{MDE}}$. This is due to the fact that $\mathcal{L}_{\text{MIM}}$ allows to model similarity of each multi-modal RS image pair with respect to other pairs. However, when $\mathcal{L}_{\text{MDE}}$ is utilized,~\basemae{s} model the similarity of each pair independently of the other pairs. Accordingly,~\basemae{s} with $\mathcal{L}_{\text{MIM}}$ achieve latent similarity preservation more accurately than those with $\mathcal{L}_{\text{MDE}}$. To this end, inter-modal latent similarity preservation is considered in all the~\basemae~models by using $\mathcal{L}_{\text{MIM}}$ for the rest of the experiments.

\subsection{Comparison with Other Approaches}
In this subsection, we analyze the effectiveness of the \mbox{\basemae} models compared to other approaches. In the first set of trials, we compare them with: 1) MAE \mbox{\cite{He:2022}}; 2) MAE-RVSA \mbox{\cite{Wang:2023}}; 3) SS-CMIR \mbox{\cite{Sumbul:2022}}; and 4) MaskVLM \mbox{\cite{Kwon:2023}}. Table \ref{tab:comparison} shows the corresponding $F_1$ scores and the required number of model parameters when the training sets of BEN-14K and BEN-270K are considered. By assessing the table, one can see that the {\basemae} models achieve the highest $F_1$ scores on almost all the CBIR tasks and training sets. As an example, when the models are trained on BEN-14K,~\basemae-CESD achieves more than 41\% higher $F_1$ score on \textit{S2} $\rightarrow$ \textit{S1} task compared to MAE. While the \mbox{\basemae} models outperform the other approaches, each \mbox{\basemae} model requires a smaller number of model parameters compared to other approaches. As an example, for both training sets, the \mbox{\basemae}-CECD leads to about 2\% higher {$F_1$} score on the \mbox{\textit{S1}} {$\rightarrow$} \mbox{\textit{S1}} task with more than 100M less model parameters compared to MaskVLM, which is one of the state-of-the-art approaches for cross-modal representation learning. Such a reduction in the number of model parameters significantly reduces the computational complexity, and thus the total training time. During inference, \basemae models require the same number of required floating point operations per second (FLOPS) compared to MAE, and thus does not increase the inference time. In greater details, when the considered training set size is increased by using BEN-270K, all the models are capable of increasing their performance. However, most of the $F_1$ scores obtained by MAE and MAE-RVSA, SS-CMIR and MaskVLM trained on BEN-270K are even lower than those obtained by the {\basemae} models trained on BEN-14K. Only MAE is capable of achieving slightly higher $F_1$ score on \textit{S2} $\rightarrow$ \textit{S2} task. These results show the effectiveness of the \mbox{\basemae} models compared to the other approaches in terms of both: i) sensor-agnostic CBIR performance; and ii) efficiency associated to the computational cost and the number of training samples.

Fig. \ref{fig:vis_res} shows an example of \textit{S2} images retrieved by all the considered models trained on BEN-270K for \textit{S1} $\rightarrow$ \textit{S2} task when the \textit{S1} query image contains the classes of \textit{Urban fabric}, \textit{Arable land} and \textit{Complex cultivation patterns}. The retrieval orders of images are given above the figure. By assessing the figure, one can observe that the~\basemae~models lead to the retrieval of similar images at all retrieval orders (e.g., the retrieved images contain all the classes associated with the query image). However, by using other models, retrieved images do not contain some of the classes that are present in the query image. As an example, on the 5th retrieved image by MaskVLM and on the 2nd retrieved image by SS-CMIR, \textit{Arable land} and \textit{Complex cultivation patterns} classes are not prominent, respectively. MAE and MAE-RVSA are not capable of retrieving images, which contain \textit{Urban fabric} class at any retrieved order.

\begin{table}[t]
\renewcommand{\arraystretch}{1.1}
\setlength\tabcolsep{2pt}
\caption{$F_1$ Scores (\%) Obtained by \mbox{\basemae}-CECD, \mbox{\basemae}-CESD, \mbox{\basemae}-SECD and \mbox{\basemae}-SESD (Trained on BEN-270K) Compared to MAE, SimMIM, Supervised ViT and CLIP (Trained on Natural Images).}
\label{tab:comp_pretrained}
\centering
\begin{tabular}{@{}ccccccc@{}}
\hline
\vspace{-0.4cm}
\multirow{2}{*}{\thead[c]{Model\\Name}} & \multirow{2}{*}{\thead[c]{Training\\Set}} &  &&&&\\
& & \multicolumn{2}{c}{\thead[c]{Uni-Modal CBIR}} & & \multicolumn{2}{c}{\thead[c]{Cross-Modal CBIR}} \\\cline{3-4} \cline{6-7}  
& & \textit{S1} $\rightarrow$ \textit{S1} & \textit{S2} $\rightarrow$ \textit{S2} & & \textit{S1} $\rightarrow$ \textit{S2} & \textit{S2} $\rightarrow$ \textit{S1} \\ \hline
MAE~\cite{He:2022} &  IN1K & 62.32 & 69.19 & & 34.81 & 36.30 \\
SimMIM~\cite{Xie:2022} &  IN1K & 64.27 & 68.90 && 36.21 & 41.47 \\
Supervised ViT~\cite{Dosovitskiy:2021} &  IN21K & 64.28 & 71.71 && 32.25 & 43.96 \\
CLIP~\cite{Radford:2021} & YFCC100M & 65.80 & 70.07 && 27.23 & 28.41 \\
\basemae-CECD & BEN-270K & 71.22 & 72.98 & & 71.61 & 72.18 \\ 
\basemae-CESD & BEN-270K & 71.00 & 72.88 & & 71.46 & 72.09 \\ 
\basemae-SECD & BEN-270K & 71.24 & 72.92 & & \textbf{71.86} & 71.94 \\ 
\basemae-SESD & BEN-270K& \textbf{71.38} & \textbf{73.16} & & 71.77 & \textbf{72.20}\\ 
\hline
\end{tabular}
\end{table}
In the second set of trials, we analyze the effectiveness of the \mbox{\basemae} models (trained on BEN-270K) compared to: 1) MAE \mbox{\cite{He:2022}}; 2) SimMIM \mbox{\cite{Xie:2022}}; 3) Supervised ViT \mbox{\cite{Dosovitskiy:2021}}; and 4) CLIP \mbox{\cite{Radford:2021}}, which are pretrained on natural images from ImageNet and YFCC100M datasets. Table \mbox{\ref{tab:comp_pretrained}} shows the corresponding $F_1$ scores. By assessing the table, one can see that the \mbox{\basemae} models achieve the highest F1 scores on both uni-modal and cross-modal CBIR tasks. As an example, \mbox{\basemae}-SESD achieves more than 7\% higher $F_1$ score on \textit{S1} $\rightarrow$ \textit{S1} task and almost 40\% higher $F_1$ score on \textit{S1} $\rightarrow$ \textit{S2} task compared to the Supervised ViT pretrained on ImageNet-21K. One can also see from the table that the effectiveness of the \mbox{\basemae} models is more evident for cross-modal CBIR tasks. While the \mbox{\basemae} models provide around 4\% higher $F_1$ score on average for uni-modal CBIR tasks, they achieve 44\% higher $F_1$ score on average for cross-modal CBIR tasks compared to the CLIP model, which is trained on 100M natural images from the YFCC100M dataset. Although the CLIP model is trained on a significantly larger training set than BEN-270K (more than two orders of magnitude), our models are capable of achieving uni-modal and cross-modal CBIR tasks more accurately than the CLIP model. These results not only show the effectiveness of our models compared to the models pretrained on computer vision images, but also the importance of self-supervised pretraining with remote sensing images.

\section{Conclusion}
\label{sec:conclusion}
In this paper, we have explored the effectiveness of MAEs for sensor-agnostic CBIR problems as a first time in RS. To this end, we have provided a systematic overview on the possible adaptations of the vanilla MAE to facilitate masked image modeling on multi-sensor RS image archives. Based on our adaptations mainly on image masking, ViT architecture and masked image modeling applied to the vanilla MAE, we have proposed different models of cross-sensor masked autoencoders (\basemae{s}). We have also conducted extensive experiments to present the sensitivity analysis and the ablation study of these~\basemae~models, while comparing them with other well-known approaches. Experimental results show the effectiveness of~\basemae{s} over other approaches in the context of sensor-agnostic CBIR in RS. Their success relies on: i) accurate transformation of the pair of encoder and decoder in the vanilla MAE into sensor-specific/common encoders, cross-sensor encoder and sensor-specific/common decoders; and ii) effective adaptation of masked image modeling by utilizing not only uni-modal reconstruction but also cross-modal reconstruction and latent similarity preservation. Based on our analyses, we have also derived
a guideline to select~\basemae~models for sensor-agnostic CBIR problems in RS as follows:
\begin{itemize}
    \item If abundant unlabeled data is available for training \basemae{s} with limited computational power, the~\basemae-CECD model can be selected. This is due to the fact that it achieves comparable results with other models, while containing significantly less number of model parameters.
    \item If abundant unlabeled data is available with enough computational sources for training \basemae{s},~\basemae-SESD can be selected since it is the best performing model among others due to its model capacity. 
    \item If the number of training samples is limited, the higher the number of parameters a~\basemae~model contains the better it performs. Thus, the selection of the~\basemae~models can be decided based on the available computational power for training.
\end{itemize}

We would like to point out that, in this paper \mbox{\basemae{s}} are studied in the context of CBIR in RS. However, \mbox{\basemae}{s} can be also considered as foundation models if they are pre-trained on larger training sets with deeper ViT backbones. Accordingly, they can be also utilized for other learning tasks as in the vanilla MAE. This can be achieved by fine-tuning the pre-trained ViTs of \mbox{\basemae}{s} on annotated RS images for various downstream tasks. Accordingly, as a future development of this work, we plan to investigate the effectiveness of the \mbox{\basemae} models for various RS learning tasks such as semantic segmentation, scene classification, etc.

\section*{Acknowledgment}
This work is supported by the European Research Council (ERC) through the ERC-2017-STG BigEarth Project under Grant 759764 and by the European Space Agency (ESA) through the Demonstrator Precursor Digital Assistant Interface For Digital Twin Earth (DA4DTE) Project. The Authors would like to thank Tom Burgert for the discussions on self-supervised learning and Dr. Yeti G\"urb\"uz for the discussions on the design of~\basemae{s}.
\bibliographystyle{IEEEtranDOI}
\bibliography{refs/defs.bib,refs/refs.bib}
\begin{IEEEbiography}
[{\includegraphics[width=1in,height=1.25in,clip,keepaspectratio]{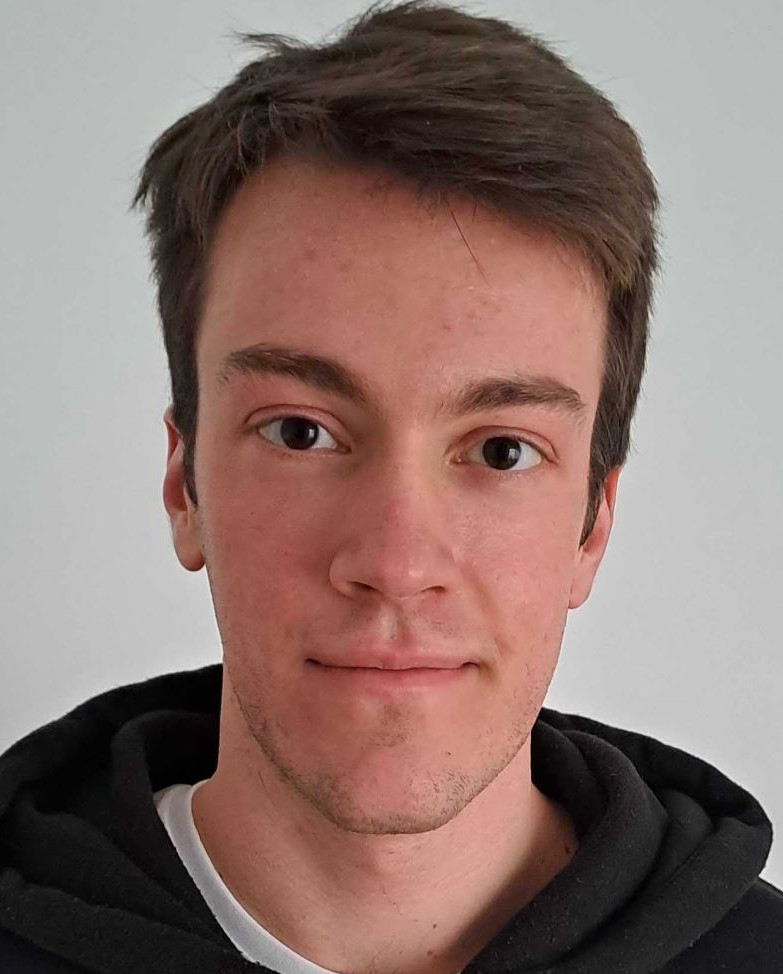}}]{Jakob Hackstein} received his B.Sc. degree in Digital Media at the University of Bremen, Germany in 2021. He is currently an M.Sc. student in Computer Science at the Technische Universität Berlin, Germany. His current research interests include deep learning and computer vision.
\end{IEEEbiography} 
\begin{IEEEbiography}[{\includegraphics[width=1in,height=1.25in,clip,keepaspectratio]{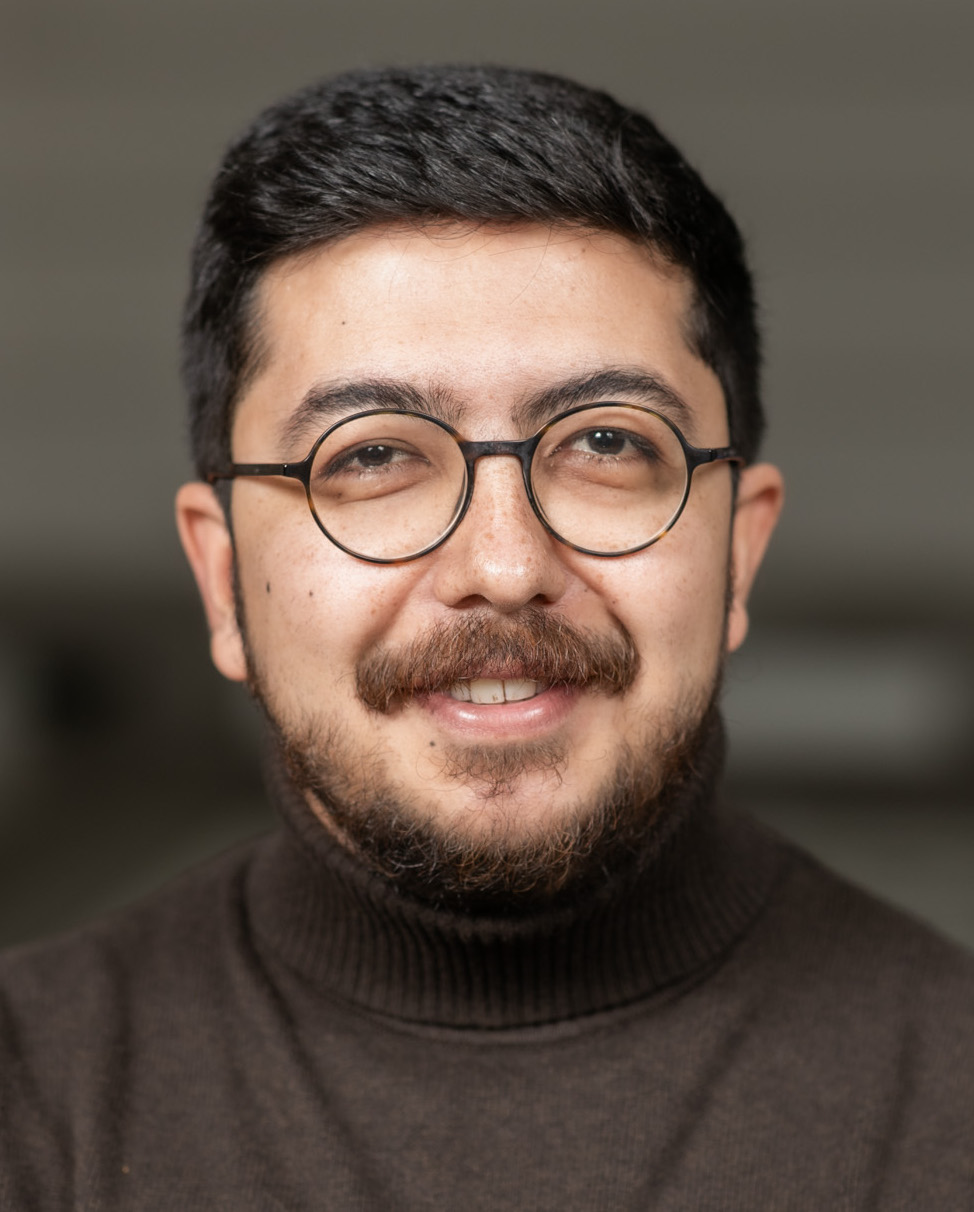}}]{Gencer Sumbul} received his B.Sc. degree in computer engineering from Bilkent University, Ankara, Turkey in 2015, the M.Sc. degree in computer engineering from Bilkent University in 2018, and the Ph.D. degree from the Faculty of Electrical Engineering and Computer Science, Technische Universit\"at Berlin, Germany in 2023. He is currently a postdoctoral scientist in the Environmental Computational Science and Earth Observation Laboratory (ECEO), École Polytechnique Fédérale de Lausanne (EPFL). His research interests include computer vision, pattern recognition and machine learning, with special interest in deep learning, large-scale image understanding and remote sensing. He is a referee for journals such as the IEEE Transactions on Image Processing, IEEE Transactions on Geoscience and Remote Sensing, the IEEE Access, the IEEE Geoscience and Remote Sensing Letters, the ISPRS Journal of Photogrammetry and Remote Sensing and international conferences such as European Conference on Computer Vision and IEEE International Geoscience and Remote Sensing Symposium.
\end{IEEEbiography} 
\vfill
\begin{IEEEbiography}[{\includegraphics[width=1in,height=1.25in,clip,keepaspectratio]{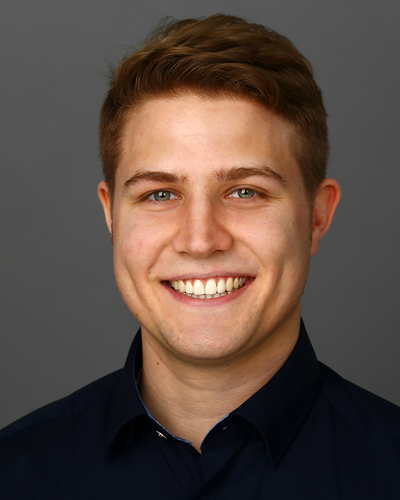}}]{Kai Norman Clasen} received his B.\,Sc.\ and M.\,Sc.\ degrees in computer engineering from Technische Universität Berlin, Berlin, Germany in 2018 and 2020, respectively. He is currently pursuing a Ph.\,D.\ degree in the Remote Sensing Image Analysis (RSiM) group at the Faculty of Electrical Engineering and Computer Science, TU Berlin and the Big Data Analytics for Earth Observation research group at the Berlin Institute for the Foundations of Learning and Data (BIFOLD). His research interests revolve around the intersection of remote sensing and deep learning, and he has a particular interest in reproducibility and open science. He is a referee for the ISPRS Photogrammetry and Remote Sensing journal.
\end{IEEEbiography} 
\begin{IEEEbiography}[{\includegraphics[width=1in,height=1.25in,clip,keepaspectratio]{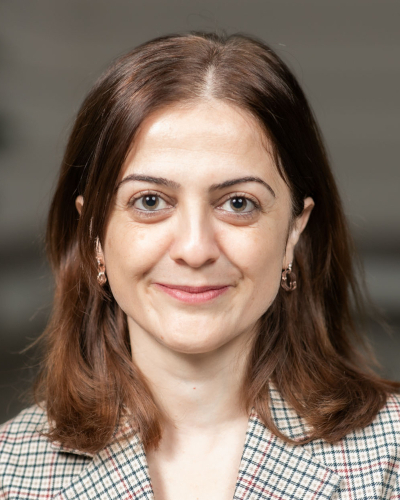}}]{Beg\"{u}m Demir} (S'06-M'11-SM'16) received the B.Sc., M.Sc., and Ph.D. degrees in electronic and telecommunication engineering from Kocaeli University, Kocaeli, Turkey, in 2005, 2007, and 2010, respectively.

She is currently a Full Professor and the founder head of the Remote Sensing Image Analysis (RSiM) group at the Faculty of Electrical Engineering and Computer Science, TU Berlin and the head of the Big Data Analytics for Earth Observation research group at the Berlin Institute for the Foundations of Learning and Data (BIFOLD). Her research activities lie at the intersection of machine learning, remote sensing and signal processing. Specifically, she performs research in the field of processing and analysis of large-scale Earth observation data acquired by airborne and satellite-borne systems. She was awarded by the prestigious ‘2018 Early Career Award’ by the IEEE Geoscience and Remote Sensing Society for her research contributions in machine learning for information retrieval in remote sensing. In 2018, she received a Starting Grant from the European Research Council (ERC) for her project “BigEarth: Accurate and Scalable Processing of Big Data in Earth Observation”. She is an IEEE Senior Member and Fellow of European Lab for Learning and Intelligent Systems (ELLIS).

Dr. Demir is a Scientific Committee member of several international conferences and workshops, such as: Conference on Content-Based Multimedia Indexing, Conference on Big Data from Space, Living Planet Symposium, International Joint Urban Remote Sensing Event, SPIE International Conference on Signal and Image Processing for Remote Sensing, Machine Learning for Earth Observation Workshop organized within the ECML/PKDD. She is a referee for several journals such as the PROCEEDINGS OF THE IEEE, the IEEE TRANSACTIONS ON GEOSCIENCE AND REMOTE SENSING, the IEEE GEOSCIENCE AND REMOTE SENSING LETTERS, the IEEE TRANSACTIONS ON IMAGE PROCESSING, Pattern Recognition, the IEEE TRANSACTIONS ON CIRCUITS AND SYSTEMS FOR VIDEO TECHNOLOGY, the IEEE JOURNAL OF SELECTED TOPICS IN SIGNAL PROCESSING, the International Journal of Remote Sensing), and several international conferences. Currently she is an Associate Editor for the IEEE GEOSCIENCE AND REMOTE SENSING LETTERS, MDPI Remote Sensing and International Journal of Remote Sensing.
\end{IEEEbiography}
\vfill
\end{document}